\documentstyle[a4,12pt,epsfig]{article}
\voffset     = - 1.20 in
\hoffset     = - 0.30 in
\begin{document}

\begin{center}
{\large EUROPEAN ORGANIZATION FOR NUCLEAR RESEARCH}
\end{center}

\bigskip

\hfill
{
\begin{tabular}{r}
CERN-EP/99-\mbox{~~~~} \\
\today                 \\
\end{tabular}
}

\bigskip

\begin{center}
{\huge\bf 
Hadronic Shower Development in Iron-Scintillator Tile Calorimetry \\
}
\end{center}

\bigskip

\begin{center}
{\large\it Submitted to Nucl.\ Instr.\ \& Meth.}
\end{center}

\bigskip

\begin{center}
{
P.~Amaral$^{k1,k2}$,
A.~Amorim$^{k1,k2}$,
K.~Anderson$^f$,
G.~Barreira$^{k1}$,
R.~Benetta$^j$,
S.~Berglund$^r$,
C.~Biscarat$^g$,
G.~Blanchot$^c$,
E.~Blucher$^f$,
A.~Bogush$^l$
% ...
}
\end{center}

{\it
\noindent
\begin{tabular}{@{}lp{6.15in}@{}}
$^a$ &  Argonne National Laboratory, Argonne, Illinois,  USA            \\
$^b$ &  University of Texas at Arlington, Arlington, Texas, USA         \\
$^c$ &  Institut de Fisica d'Altes Energies, 
        Universitat Aut\`onoma de Barcelona, 
        
        Barcelona, Spain                                                \\
$^d$ &  Comenius University, Bratislava, Slovakia                       \\
$^e$ &  Institute of Atomic Physics, Bucharest, Rumania                 \\
$^f$ &  University of Chicago, Chicago, Illinois, USA                   \\
$^g$ &  LPC Clermont--Ferrand, Universit\'e Blaise Pascal / CNRS--IN2P3, 

        Clermont--Ferrand, France                                       \\
$^h$ &  JINR, Dubna, Russia                                             \\
$^i$ &  Michigan State University, East Lansin, Michigan, USA           \\
$^j$ &  CERN, Geneva, Switzerland                                       \\
$^k$ &  1) LIP Lisbon, 
        2) FCUL Univ.\ of Lisbon, 
        3) LIP and FCTUC Univ.\ of Coimbra,

        4) Univ.\ Cat\'olica Figueira da Foz, Portugal                  \\
$^l$ &  Institute of Physics, National Academy of Science, 
        Minsk, Republic of Belarus                                      \\
$^m$ &  Pisa University and INFN, Pisa, Italy                           \\
$^n$ &  Charles University, Prague, Czech Republic                      \\
$^o$ &  Academy of Science, Prague, Czech Republic                      \\
$^p$ &  Institute for High Energy Physics, Protvino, Russia             \\
$^q$ &  COPPE/EE/UFRJ, Rio de Janeiro, Brazil                           \\
$^r$ &  Stockholm University, Stockholm, Sweden                         \\
$^s$ &  University of Illinois, Urbana--Champaign, Illinois, USA        \\
$^t$ &  IFIC, Centro Mixto Universidad de Valencia-CSIC, 
        E46100 Burjassot, Valencia, Spain                               \\
$^u$ &  Yerevan Physics Institute, Yerevan, Armenia                     \\
\end{tabular}
}

\vspace*{\fill}

\begin{abstract}
The lateral and  longitudinal profiles of hadronic showers detected by
a prototype of the ATLAS Iron-Scintillator Tile Hadron Calorimeter
have been investigated.
This calorimeter uses a unique longitudinal configuration of 
scintillator tiles.
Using a fine-grained pion beam scan at 100 GeV, a detailed picture of  
transverse shower behavior is obtained.
The underlying radial energy densities for four depth segments and for 
the entire calorimeter have been reconstructed.
A three-dimensional hadronic shower parametrization has been developed.
The results presented here are useful for understanding the performance 
of iron-scintillator calorimeters, 
for developing fast simulations of hadronic showers, 
for many calorimetry problems requiring the integration of a shower 
energy deposition in a volume and for future calorimeter design.  
\end{abstract}

\newpage

\section{Introduction}

\begin{figure*}[tbph]
     \begin{center}
        \begin{tabular}{|c|}
        \hline
\mbox{\epsfig{figure=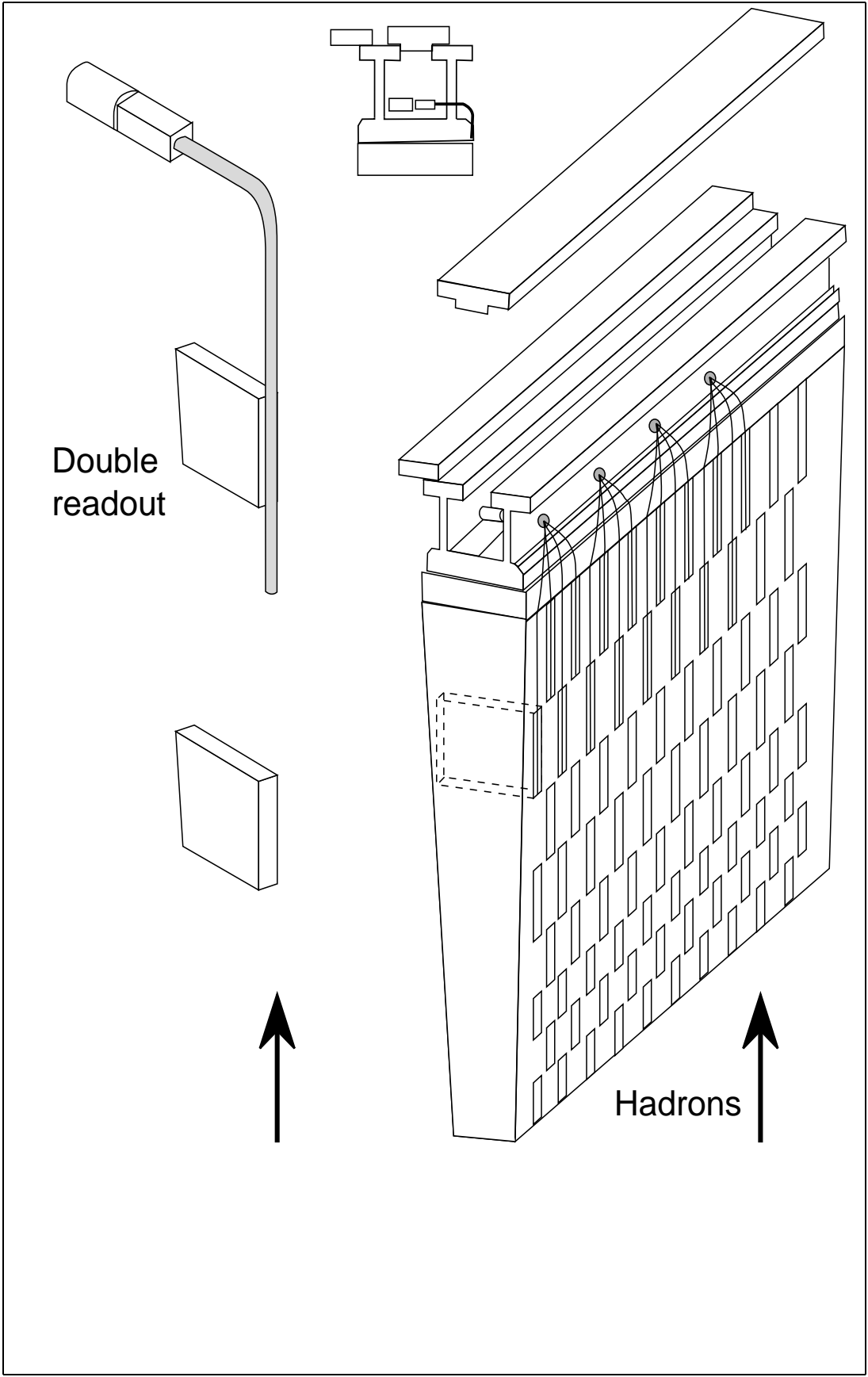,width=0.85\textwidth,height=0.9\textheight}}
\\
        \hline
        \end{tabular}
     \end{center}
       \caption{
       Conceptual design of a Tile calorimeter module.
       \label{fig:f1}}
\end{figure*}

We report on an experimental study of hadronic shower profiles detected by
the prototype of the ATLAS Barrel Tile Hadron Calorimeter (Tile calorimeter) 
\cite{atcol94}, \cite{tilecal-tdr96}.
The innovative design of this calorimeter, using longitudinal segmentation 
of active and passive layers (see Fig.\ \ref{fig:f1}), provides an 
interesting system for the measurement of hadronic shower profiles.
Specifically, we have studied the transverse development of hadronic showers 
using 100 GeV pion beams and longitudinal development of hadronic showers 
using 20 -- 300 GeV pion beams.

Characteristics of shower development in hadron calorimeters have been
published for some time.  
However, a complete quantitative description of transverse 
and longitudinal properties of hadronic showers does not exist 
\cite{bock97}.
The transverse profiles are usually expressed as a 
function of transverse coordinates, not the radius, 
and are integrated over the other coordinate
\cite{womersley88}.
The three-dimensional parametrization of hadronic showers described here 
could be a useful starting point for fast simulations,
which can be faster than full simulations at the microscopic level 
by several orders of magnitude  
\cite{bock81}, \cite{grindhammer90}, \cite{brun91}.

The paper is organised as follows.
In Section 2, the calorimeter and the test beam setup are briefly described.
In Section 3, the mathematical procedures for extracting the underling 
radial energy density of hadronic showers are developed.
The obtained results on the transverse and longitudinal profiles, 
the radial energy densities and the radial containment  
of hadronic shower are presented in Sections 4 -- 7.
Section 8 and 9 investigate the three-dimensional parametrization 
and electromagnetic fraction of hadronic shower.
Finally Section 10 contains a summary and the conclusions.

\section{The Calorimeter}

\begin{figure*}[tbph]
     \begin{center}
        \begin{tabular}{|c|}
        \hline
\mbox{\epsfig{figure=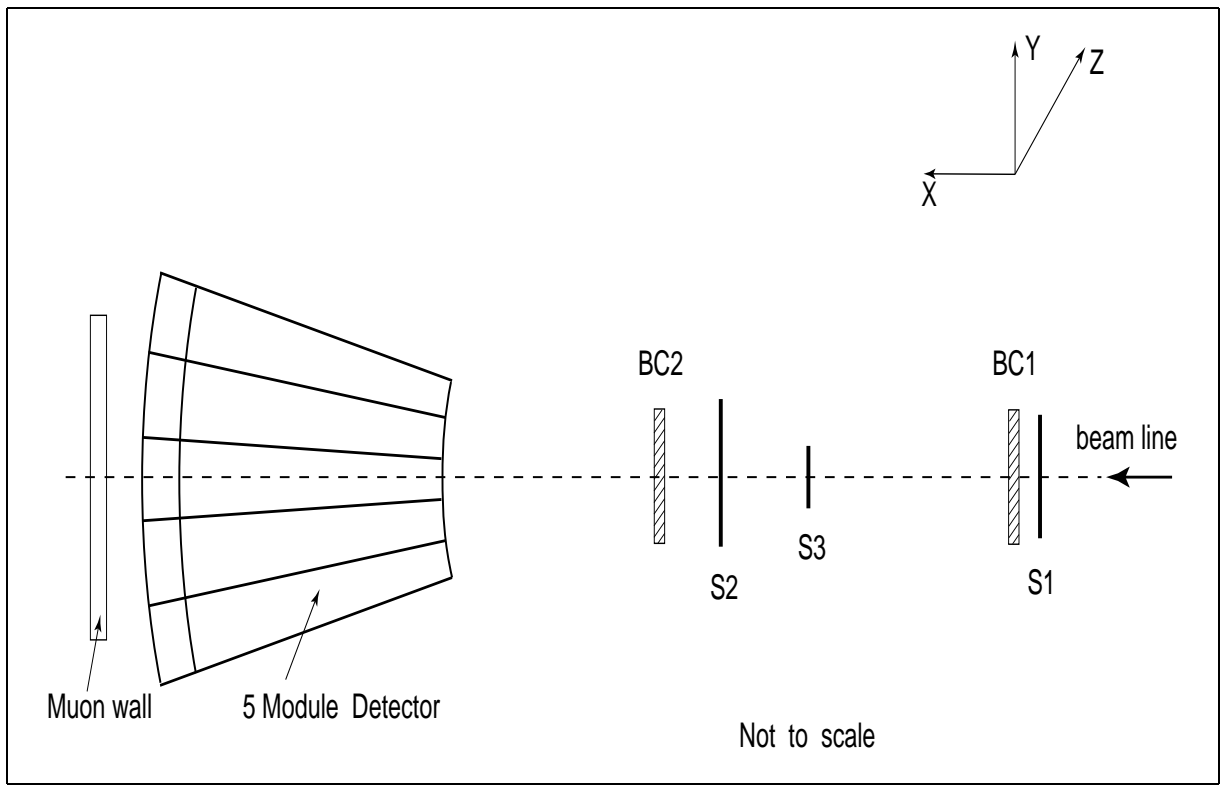,width=0.95\textwidth,height=0.4\textheight}}
\\
        \hline
        \end{tabular}
     \end{center}
       \caption{
       Schematic layout of the experimental setup.
       S1 -- S3 are beam trigger scintillators, and BC1 -- BC2 are 
      (Z,Y) proportional chambers.
       \label{fig:f01}}
\end{figure*}

The prototype Tile Calorimeter used for this study is composed 
of five modules 
stacked in the $Y$ direction, as shown in Fig.\ \ref{fig:f01}.
Each module spans $2 \pi / 64$ in the azimuthal angle, 
100 cm in the $Z$ direction,
180 cm in the $X$ direction (about 9 interaction lengths, 
$\lambda_I$, or about 80 
effective radiation lengths, $X_{0}$), and has a front face of 100 
$\times$ 20 cm$^2$
\cite{berger95}.
The absorber structure of each module consists of 57 repeated ``periods".
Each period is 18 mm thick and consists of four layers.
The first and third layers are formed by large trapezoidal steel plates
(master plates) and span the full longitudinal dimension of the module.
In the second and fourth layers, smaller trapezoidal steel plates
(spacer plates) and scintillator tiles alternate along the $X$ direction.
These layers consist of 18 different trapezoids of steel or scintillator,
each of 100 mm in depth.
The master plates, spacer plates and scintillator tiles are 
5 mm, 4 mm and 3 mm thick, respectively.
The iron to scintillator ratio is $4.67 : 1$ by volume.
The calorimeter thickness along the beam direction at the incidence angle
(the angle between the incident particle direction and the normal to the 
calorimeter front face) of $\Theta = 10^{\circ}$ corresponds to 1.49 m 
of iron equivalent
\cite{lokajicek95-64}.

Wavelength shifting fibers collect scintillation light from the 
tiles at both of their open (azimuthal) edges and bring it to 
photo-multipliers 
(PMTs) at the periphery of the calorimeter.
Each PMT views a specific group of tiles through the corresponding 
bundle of fibers.
The modules are divided into five segments along $Z$.
They are also longitudinally segmented (along $X$) into four depth segments.
The readout cells have a lateral dimensions of 200 mm along $Z$, 
and longitudinal 
dimensions of 300, 400, 500, 600 mm for depth segments 1 -- 4, corresponding 
to 1.5, 2, 
2.5 and 3 $\lambda_{I}$ at $\Theta = 0^{\circ}$ respectively.
Along $Y$, the cell sizes vary between about 200 and 370 mm depending on
the X (depth) coordinate. 
We recorded energies for 100 different cells for each event 
\cite{berger95}.

The calorimeter was placed on a scanning table that allowed
movement in any direction.
Upstream of the calorimeter, a trigger counter telescope 
(S1 -- S3) was installed,
defining a beam spot approximately 20 mm in diameter.
Two delay-line wire chambers (BC1 -- BC2),  
each with $(Z, Y)$  readout,
allowed the impact point of beam particles on the
calorimeter face to be reconstructed to better than $\pm 1$ mm
\cite{ariztizabal94}.
``Muon walls'' were placed behind (800$\times$800 mm$^2$) 
and on the positive $Z$ side (400$\times$1150 mm$^2$) of the calorimeter
modules
to measure longitudinal and lateral hadronic shower leakage
\cite{lokajicek95-63}.

We used the TILEMON program \cite{efthymiopoulos95} to convert the 
raw calorimeter data into PAW Ntuples \cite{paw} containing 
calibrated cell energies and other information used in this study.  

The data used for the study of lateral profiles were collected 
in 1995 during 
a special $Z$-scan run at the CERN SPS test beam.  
The calorimeter was exposed to 100 GeV negative pions at a $10^{\circ}$ 
angle with varying impact points in the $Z$-range from $- 360$ to $+ 200$ mm.
A total of $>$~300,000 events have been analysed;
for the lateral profile study only events without lateral leakage were
used.
The uniformity of the calorimeter's response for this $Z$-scan is
estimated 
to be $1\%$  
\cite{budagov96-76}.

The data used for the study of longitudinal profiles were obtained using 
20 -- 300 GeV negative pions at a $20^{\circ}$ angle and were also taken
in 1995
during the same test beam run. 

\section{Extracting the Underlying Radial Energy Density}

In this investigation we use a coordinate system based on the incident 
particle direction. 
The impact point of the incident particle at the calorimeter front face 
defines the origin of coordinate system.
The incident particle direction forms the $x$ axis, 
while the $y$ axis is in the same direction as $Y$ defined in Section 2. 
The normal to the $xy$ surface 
defines 
the $z$ axis. 

We measure the energy deposition in each calorimeter cell for every event.
In the $ijk$-cell of the calorimeter with the volume $V_{ijk}$ and cell 
center coordinates $(x_{c},y_{c},z_{c})$, the energy deposition $E_{ijk}$ is
\begin{equation}
        E_{ijk}(x_{c},y_{c},z_{c}) = 
                \int \limits_{V_{ijk}} f(x,y,z)\ dx dy dz,
\label{e010}
\end{equation}
where  $f(x,y,z)$ is the three-dimensional hadronic shower 
energy density function.
Due to the azimuthal symmetry of shower profiles, the density $f(x,y,z)$ is 
only a function of the radius $r = \sqrt{y^2+z^2}$ from the shower axis
and the longitudinal coordinate $x$.
Then 
\begin{equation}
        E_{ijk}(x_{c},y_{c},z_{c}) = 
                \int \limits_{V_{ijk}} \Psi(x, r)\ rdr d \phi dx,
\label{e010-01}
\end{equation}
where $\phi$ is the azimuthal angle and  
$\Psi(x,r)$ has the form of a joint probability density function
(p.d.f.) 
\cite{review96}. 
The joint p.d.f. can be further decomposed as a product of 
the marginal p.d.f., $dE  (x)/ dx$, 
and the conditional p.d.f., $\Phi (x, r)$, 
\begin{equation}
        \Psi(x,r) = 
                \frac{dE(x)}{dx} \cdot \Phi (x,r). 
\label{e123-10}
\end{equation}
The longitudinal density $dE/dx$ is defined as 
\begin{equation}
        \frac{dE(x)}{dx} =
            \int \limits_{-\infty}^{\infty}
            \int \limits_{-\infty}^{\infty}
            f(x,y,z)\ dy dz . 
\label{e08}
\end{equation}
Finally, the radial density function $\Phi(r)$ for a given depth segment is
\begin{equation}
        \Phi(r) = E_{0}\ \Phi(x,r) , 
\label{e123-001}
\end{equation}
where $E_{0}$ is the total shower energy deposition into the depth segment
for fixed $x$.

There are several methods for extracting the radial density $\Phi(r)$
from the measured distributions of energy depositions.
One method is to unfold $\Phi(r)$ using expression (\ref{e010-01}).
This method was used in the analysis of the data from the lead-scintillating 
fiber calorimeter \cite{acosta92}.
Several analytic forms of $\Phi(r)$  were tried,
but the simplest that describes the 
energy deposition in cells was a combination of an exponential 
and a Gaussian:
\begin{equation}
        \Phi(r) = 
                \frac{b_1}{r}\ e^{ - \frac{r}{\mu_1}} +
                \frac{b_2}{r}\ e^{ - {( \frac{r}{\mu_{2}} )}^2 },
\label{e1}
\end{equation}
where $b_i$ and $\mu_i$ are the free parameters.

Another method for extracting the radial density 
is to use the marginal density function
\begin{equation}
        f(z) = 
                \int \limits_{-\infty}^{\infty}  
                \int \limits_{x_1}^{x_2} 
                f(x,y,z)\ dx dy .
\label{e09-1}
\end{equation}
which is related to the radial density $\Phi(r)$
\begin{equation}
        f(z) = 
                2 \int \limits_{|z|}^{\infty} 
                \frac{\Phi(r)\ r dr}{\sqrt{r^{2}-z^{2}}} .
\label{e4}
\end{equation}
This method was used \cite{lednev95} 
for extracting the electron shower transverse profile
from the GAMS-2000 electromagnetic calorimeter data
\cite{akopdijanov77}.
The above integral equation (\ref{e4}) can be reduced to an Abelian equation
by replacing variables 
\cite{whitteker27}.
In \cite{budagov97},
the following solution to equation (\ref{e4}) was obtained 
\begin{equation}
        \Phi(r) = 
                - \frac{1}{\pi}\ \frac{d}{dr^2}
                \int \limits_{r^2}^{\infty} 
                \frac{f(z)\ d z^2}{\sqrt{z^{2}-r^{2}}}.
\label{e5}
\end{equation}

For our study, we used the sum of three exponential functions to
parameterize $f(z)$ as
\begin{equation}
        f(z)  = 
                \frac{E_{0}}{2B}\  \sum_{i=1}^{3}\  a_i\ 
                e^{ - \frac{|z|}{{\lambda}_{i}} }, 
\label{e21}
\end{equation}
where 
$z$ is the transverse coordinate,
$E_{0},\ a_i,\ \lambda_i$ are free parameters, 
$B =  \sum_{i=1}^{3} a_i \lambda_i$, 
$\sum_{i=1}^{3} a_i = 1$
and $\int_{-\infty}^{+\infty} f  (z) dz = E_{0}$.
In this case, the radial density function, obtained by integration 
and differentiation of equation (\ref{e5}), is 
\begin{equation}
        \Phi(r) = 
                \frac{E_{0}}{2 \pi B}\ 
                \sum_{i=1}^{3}\ \frac{a_{i}}{\lambda_{i}}\
                K_{0} \biggl( \frac{r}{\lambda_{i}} \biggr), 
\label{e23}
\end{equation}
where 
$K_{0}$ is the modified Bessel function.
This function goes to $\infty$\ as $r \rightarrow 0$\
and goes to zero as $r \rightarrow \infty$.

We define a column of five cells in a depth segment as a tower. 
Using the parametrization shown in equation (\ref{e21}),
we can show that the energy deposition in a tower \cite{gavrish91},
$E(z) = \int_{z - h/2}^{z + h/2} f(z) dz$, 
can be written  as
\begin{eqnarray}
        E(z)  = 
&               E_{0} - 
               \frac{E_{0}}{B}\  \sum \limits_{i=1}^{3}\  a_i {\lambda}_i\
                \cosh (\frac{ |z| }{ {\lambda}_i }) \
                e^{ - \frac{h}{2 {\lambda}_i} },  
&  for\         |z| \leq \frac{h}{2},
\\
        E(z) =  
&              \frac{E_{0}}{B}\ \sum \limits_{i = 1}^{3}\ a_i {\lambda}_i\
                \sinh (\frac{h}{2 \lambda_i }) \ 
                e^{ - \frac{|z|}{{\lambda}_i}},
&  for\         |z| > \frac{h}{2},
\label{e023}
\end{eqnarray}
where $h$ is the size of the front face of the tower along the $z$ axis.
Note that as $h \rightarrow 0$, we get $E (z) / h \rightarrow f(z)$.
As $h \rightarrow \infty $, we find that $E (0) \rightarrow E_{0}$.

The full width at half maximum (FWHM)
of an energy deposition profile for small values of $(FWHM - h)/h$  
can be approximated by  
\begin{equation}
        FWHM = h +
                \frac{2E(h/2) - E(0)}{ 
                \frac{E_{0}}{B}\
                \sum \limits_{i=1}^{3}\ a_i\ 
                \sinh ( \frac{h}{2 \lambda_i }) \
                e^{- \frac{h}{2 \lambda_i}}
                }.
\label{e0023}
\end{equation}
We will show below that this approximation agrees well with our data.

A cumulative function may be derived from the density function as
\begin{equation}
        F(z) = 
                 \int \limits_{-\infty}^{z} f(z)\ dz .
\label{e3-01}
\end{equation}
For our parametrization in equation (\ref{e21}), the cumulative
function becomes
\begin{eqnarray}
        F(z) = 
&               \frac{E_{0}}{2B}\ \sum \limits_{i=1}^{3}\  a_i\ \lambda_i\
                e^{  \frac{z}{ \lambda_{i} } },  
& for\          z \leq 0,
\\
        F(z) = 
&               E_{0} - 
                \frac{E_{0}}{2B}\ \sum \limits_{i=1}^{3}\  a_i\ \lambda_i\
                e^{- \frac{z}{\lambda_{i}}}, 
& for\          z > 0,
\label{e23-2}
\end{eqnarray}
where 
$z$ is the position of the edge of a tower along the $z$ axis.
Note that the cumulative function does not depend on the cell size 
$h$. 
We can construct the cumulative function and deconvolute
the density f(z) from it for any size calorimeter cell.
Note also that cumulative function is well behaved at the key points:
$F( - \infty ) = 0$,\ $F( 0 ) = E_{0} / 2$,\ 
and $F( \infty ) = E_{0}$.

The radial containment of a shower as a function of $r$ is given by
\begin{equation}
        I(r) = 
                \int \limits_0^r \int \limits_0^{2 \pi} 
                \Phi (r)\ r dr d\phi =
                E_{0} -
                \frac{E_{0} r}{B}\ \sum \limits_{i=1}^{3}\ a_{i}\ 
                K_1 \biggl( \frac{r}{{\lambda}_{i}} \biggr),
\label{e008}
\end{equation}
where $K_{1}$ is the modified Bessel function.
As $r \rightarrow \infty$, the function 
$r K_1 (r)$ tends to zero and we get $I(\infty) = E_{0}$, as expected. 

We use two methods to extract the radial density function 
$\Phi(r)$.
One method is to unfold $\Phi(r)$ from (\ref{e010-01}).
Another method is to use the expression (\ref{e5}) after
we have obtained the marginal density function $f(z)$.
There are three ways to extract $f(z)$:
by fitting the energy deposition $E(z)$ \cite{gavrish91},
by fitting the cumulative function $F(z)$ and
by directly extracting $f(z)$ by numerical differentiation of 
the cumulative function. 
The effectiveness of these various methods depend on the 
scope and quality of the experimental data.

\section{Transverse Behaviour of Hadronic Showers}

\begin{figure*}[tbph]
     \begin{center}
        \begin{tabular}{cc}
\mbox{\epsfig{figure=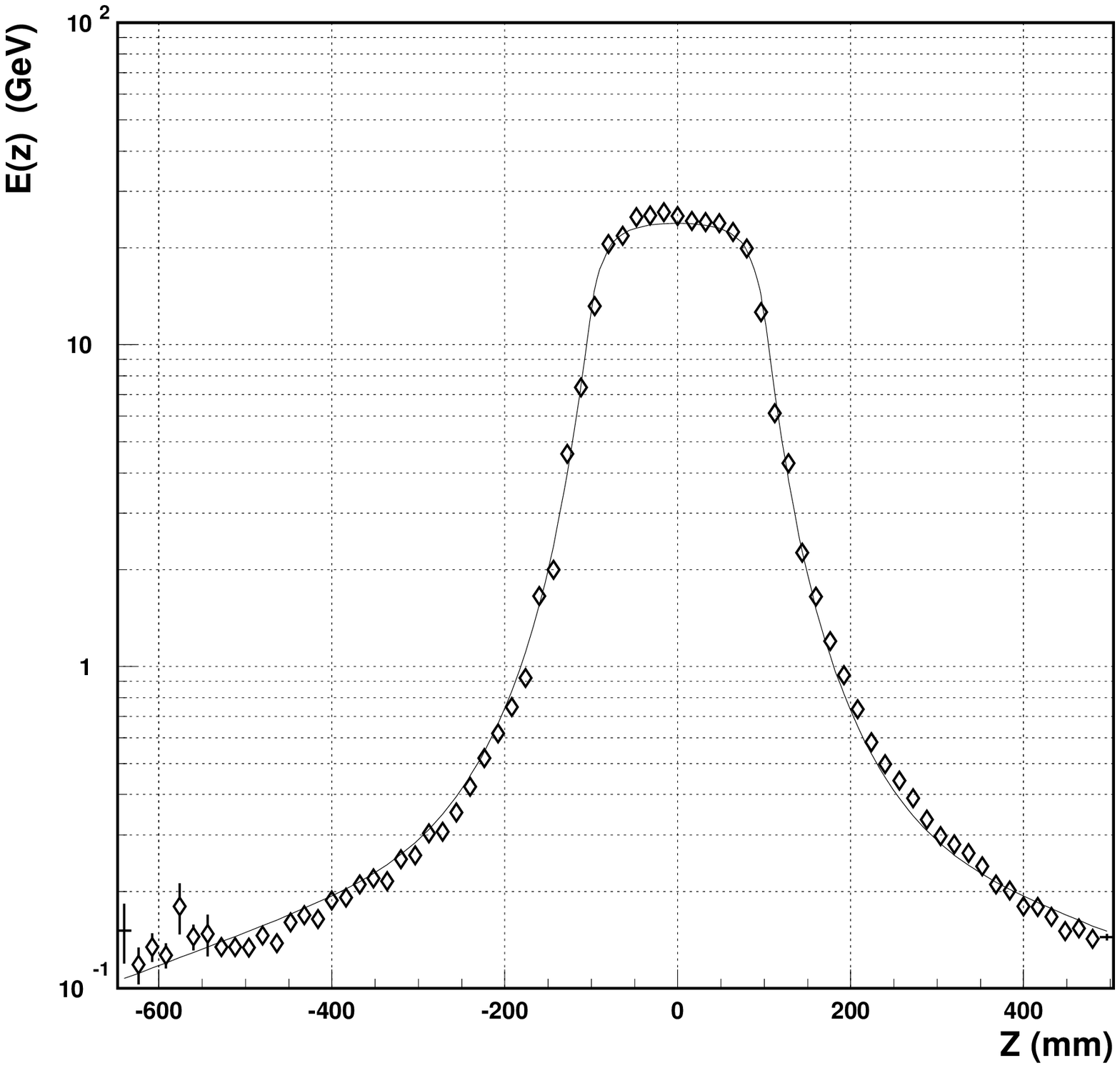,width=0.45\textwidth,height=0.4\textheight}}
&
\mbox{\epsfig{figure=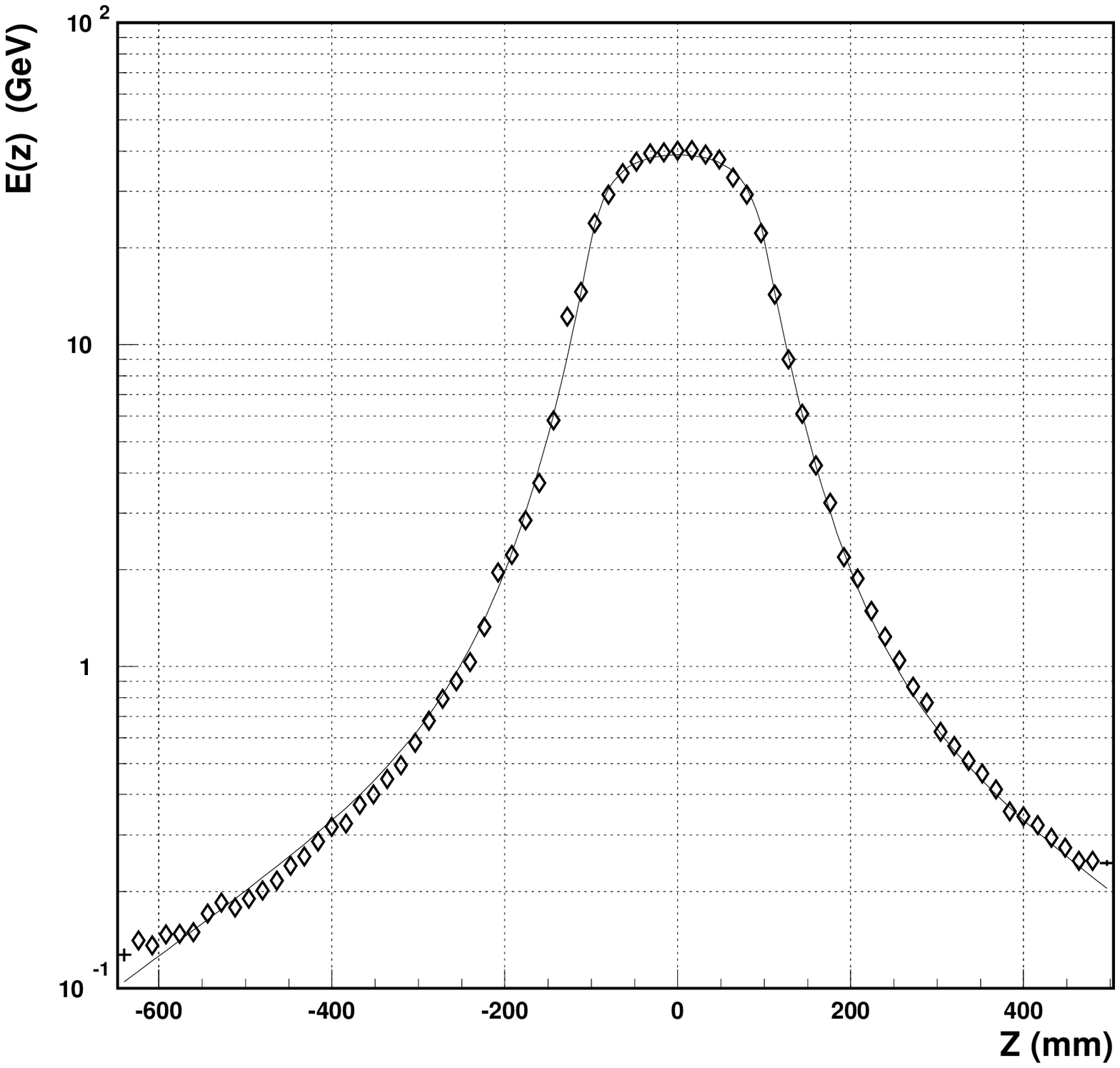,width=0.45\textwidth,height=0.4\textheight}}
        \\
\mbox{\epsfig{figure=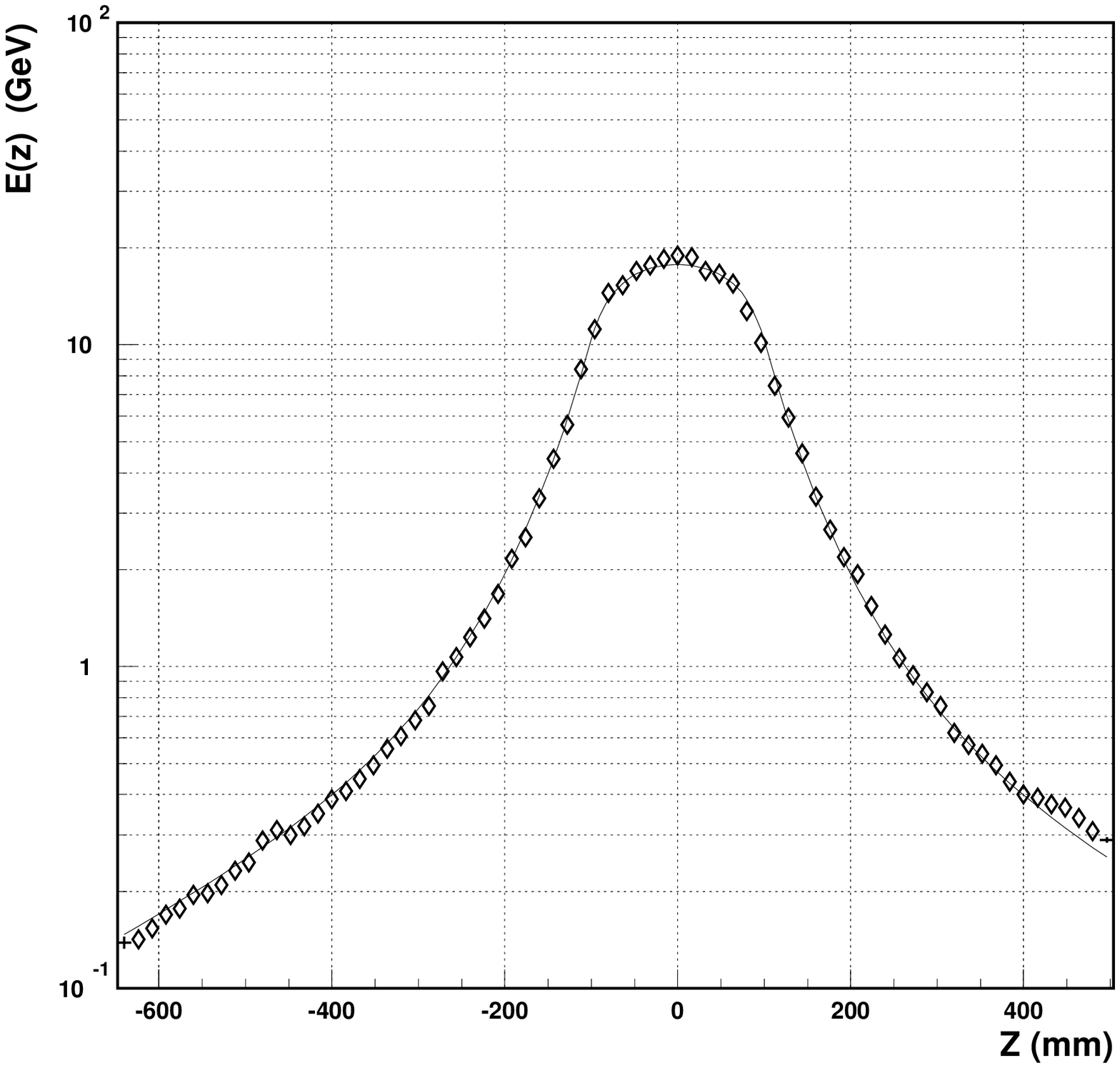,width=0.45\textwidth,height=0.4\textheight}}
&
\mbox{\epsfig{figure=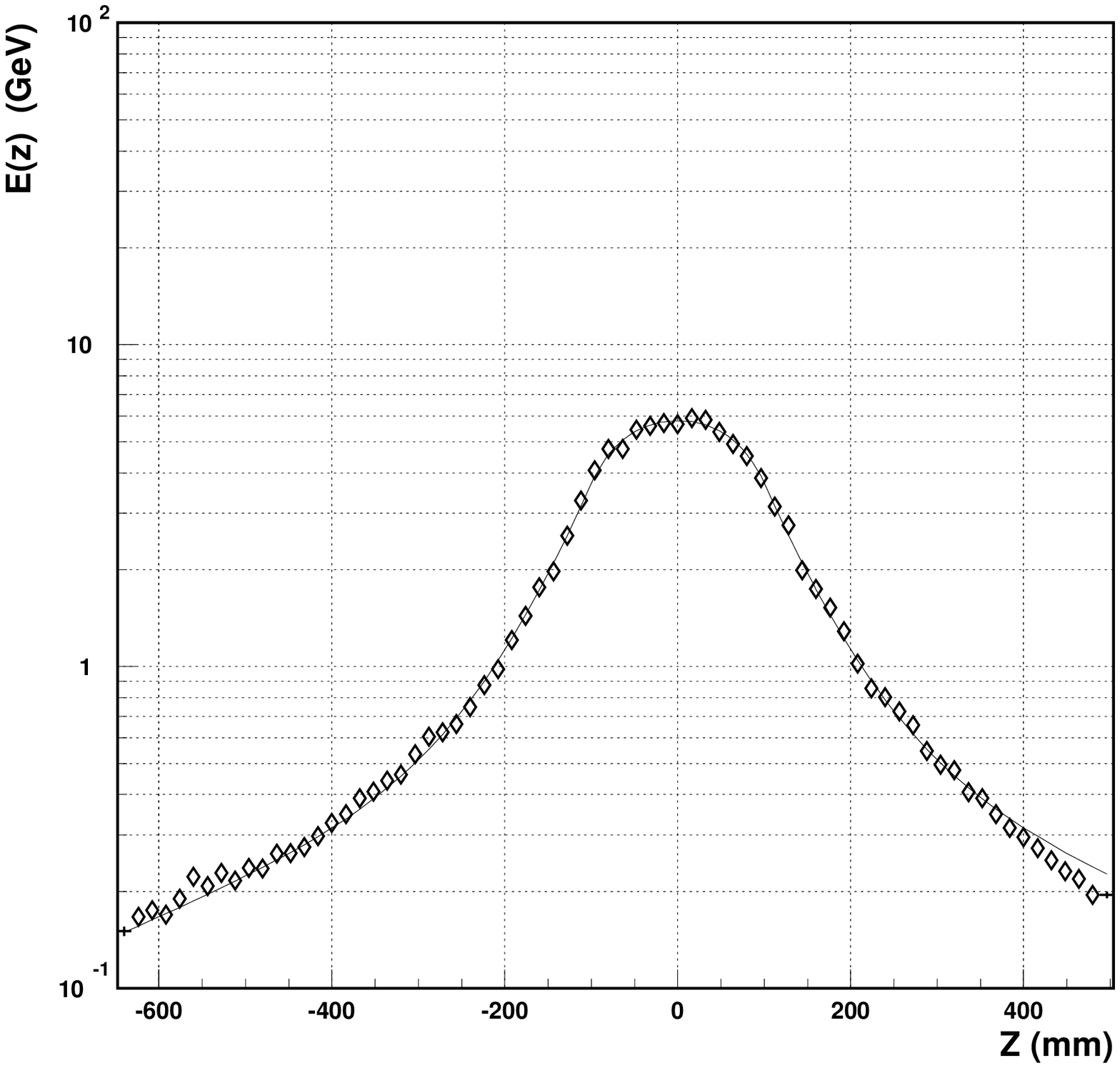,width=0.45\textwidth,height=0.4\textheight}}
        \\
        \end{tabular}
     \end{center}
      \caption{
        Energy depositions of 100 GeV pions in towers 
        of depth segments 1 -- 4  
        as a function of the $z$ coordinate:
        top left is for depth segment 1,
        top right is for depth segment 2,
        bottom left is for depth segment 3,
        bottom right is for depth segment 4. 
        Only statistical errors are shown.
        Curves are fits of equations (12) and (\ref{e023}) to the data.
       \label{fig:f5-001}}
\end{figure*}

\begin{figure*}[tbph]
     \begin{center}
        \begin{tabular}{c}
\mbox{\epsfig{figure=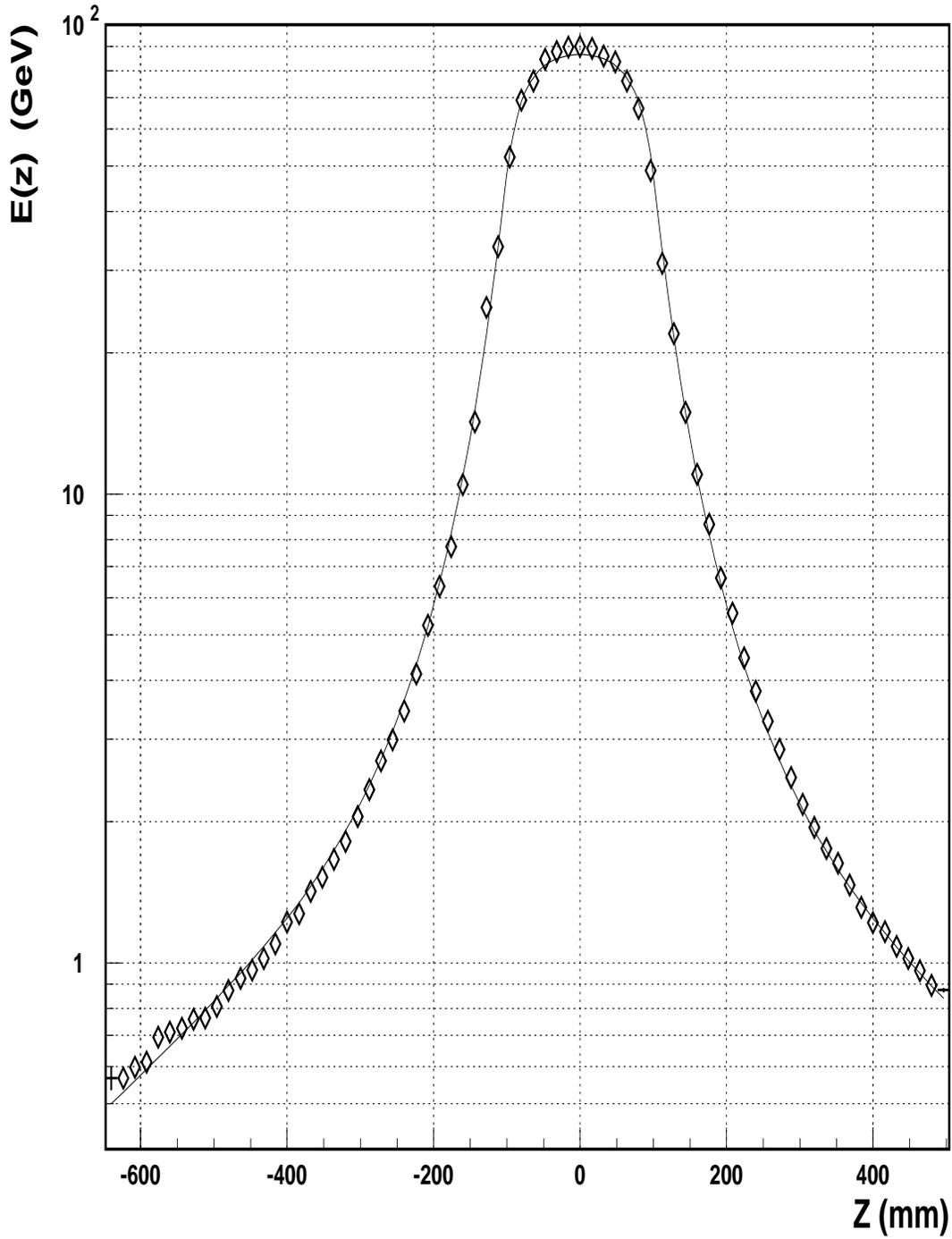,width=0.95\textwidth,height=0.85\textheight}}
        \\
        \end{tabular}
     \end{center}
      \caption{
        Energy depositions in towers,  summed over all calorimeter 
        depth segments,  as a function of the $z$ coordinate. 
        Only statistical errors are shown.
        The curve is the result of the fit by formulas (12) and (\ref{e023}).
       \label{fig:f6}}
\end{figure*}

Figure \ref{fig:f5-001} shows the energy depositions in towers  
for depth segments 1 -- 4 as a function of the $z$ coordinate of the 
center of the tower.
Figure \ref{fig:f6} shows the same for the entire calorimeter
(the sum of the histograms presented in 
Fig.\ \ref{fig:f5-001}).

\begin{figure*}[tbph]
     \begin{center}
        \begin{tabular}{c}
\mbox{\epsfig{figure=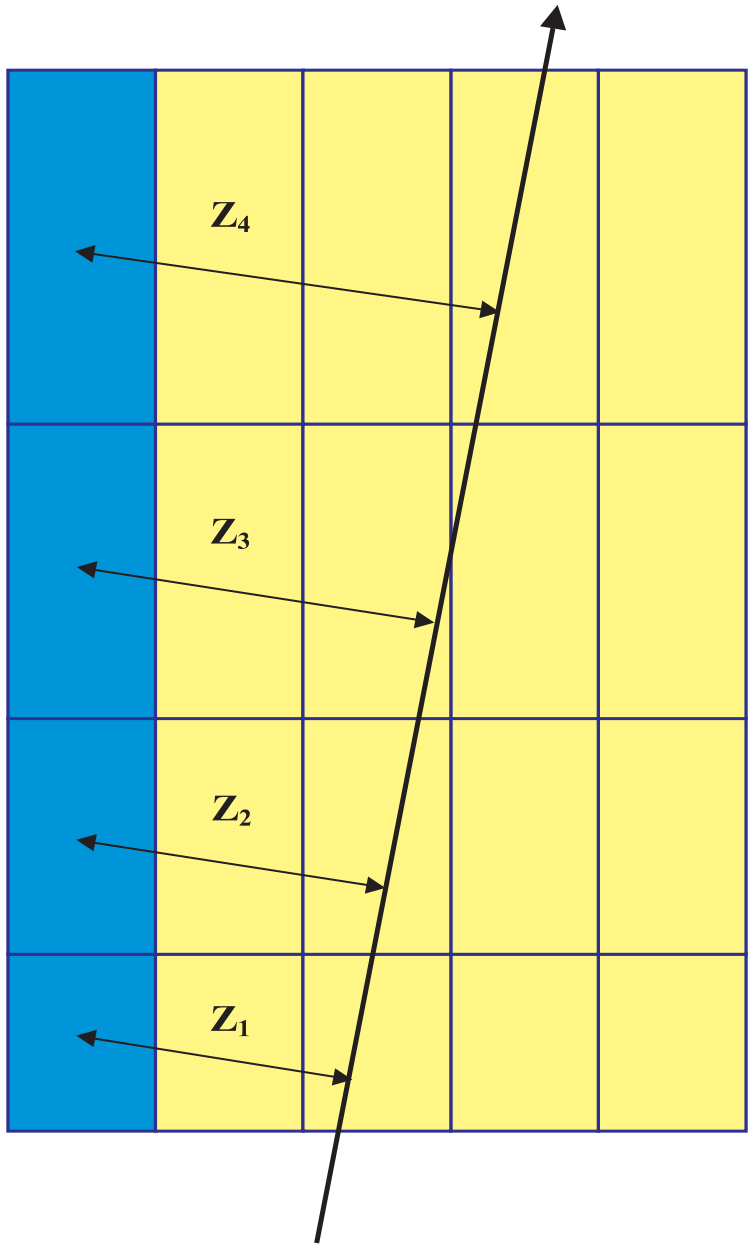,width=0.6\textwidth,height=0.65\textheight}}\\
        \end{tabular}
     \end{center}
       \caption{
       Schematic layout (top view) of Tile calorimeter 
       experimental setup.
       $z_1$ -- $z_4$ are the distances between the centre of towers 
       (for the four depth segments) and the direction of the beam
       particle.
       \label{fig:f01-1}}
\end{figure*}

These Figures are profile histograms \cite{paw} and
give the energy deposited in any tower
for all analysed events in bins of the $z$ coordinate.
Here the coordinate system is linked to the incident particle direction
where $z=0$ is the coordinate of the beam impact points at the 
calorimeter front face.
Figure \ref{fig:f01-1}
schematically shows a top view of the experimental setup and indicates the 
$z$ coordinate of each tower. 
The $z_1$, $z_2$, $z_3$ and $z_4$ 
are the distances between the centre of towers 
(for the four depth segments) and the direction of beam
particle.
The values of the $z$-coordinate of the tower centers are negative 
to the left of the beam, positive to the right of the beam and
range from $- 750$ mm to $+ 600$ mm.
To avoid edge effects, we present tower energy depositions in the range 
from $- 650$ mm to $+ 500$ mm.
Note that the total tower height 
(about 1.0 m at the front face, and about 1.8 m at the back) 
is sufficient for shower measurements without significant leakage
in the vertical direction. 

As mentioned earlier, events with significant lateral leakage
(identified by a clear minimum-ionising signal
in the lateral muon wall) were discarded. The resulting left-right
asymmetries in the distributions of Figures \ref{fig:f5-001}  
and \ref{fig:f6} are very small.

As will be shown later (Section 6), the $99\%$ containment 
radius is less than $500$ mm. 

The fine-grained $z$-scan provided many different beam
impact locations within the ca\-lorimeter.
Due to this, we obtained a detailed picture of the 
transverse shower behaviour in the calorimeter.
The tower energy depositions shown in 
Figures \ref{fig:f5-001} and \ref{fig:f6}
span a range of about three orders of magnitude.
The plateau for $|z| < 100$ mm ($h/2$) and the fall-off at large $|z|$
are apparent.
Similar behaviour of the transverse profiles was observed in other 
calorimeters as well
\cite{acosta92}, \cite{gavrish91}.

We used the distributions in Figs.\ \ref{fig:f5-001} and \ref{fig:f6}
to extract the underlying 
mar\-gi\-nal densities function for four depth segments of the calorimeter
and for the entire calorimeter.
The solid curves in these figures
are the results of the fit with equations (12) and  (\ref{e023}).
The fits typically differ from the experimental distribution by less
than $5\%$.

In comparison with 
\cite{gavrish91} and \cite{binon83},
where the transverse profiles exist only for distances less than 
$250$ mm, our more extended profiles (up to $650$ mm) require that  
the third exponential term be introduced.
The parameters $a_i$ and $\lambda_i$, obtained by fitting, are listed in 
Table~\ref{Tb2}. 
The values of the parameter $E_{0}$, the average energy shower deposition 
in a given depth segment,
are listed in Table~\ref{long-1}. 

We have compared our values of $\lambda_1$ and $\lambda_2$ 
with the ones from the conventional iron-scintillator calorimeter
described in \cite{binon83}.
At $100$ GeV, our results for the entire calorimeter are
$23\pm1$ mm and $58\pm4$ mm for $\lambda_1$ and $\lambda_2$ respectively.
They agree well with the ones from \cite{binon83},
which are $18\pm3$ mm and $57\pm4$ mm.

We determined the FWHM of energy deposition profiles 
(Figs.\ \ref{fig:f5-001} and \ref{fig:f6}) using formula (\ref{e0023}).
The characteristic FWHM are found to be approximately equal to
transverse tower size.
The relative difference of FWHM from
transverse tower size, $(FWHM - h)/h$, amount to  $2\%$
for depth segment 1, depth segment 2 and for the entire calorimeter, 
$7\%$ for depth segment 3 and $15\%$ for depth segment 4.

\begin{figure*}[tbph]
     \begin{center}
        \begin{tabular}{c}
\mbox{\epsfig{figure=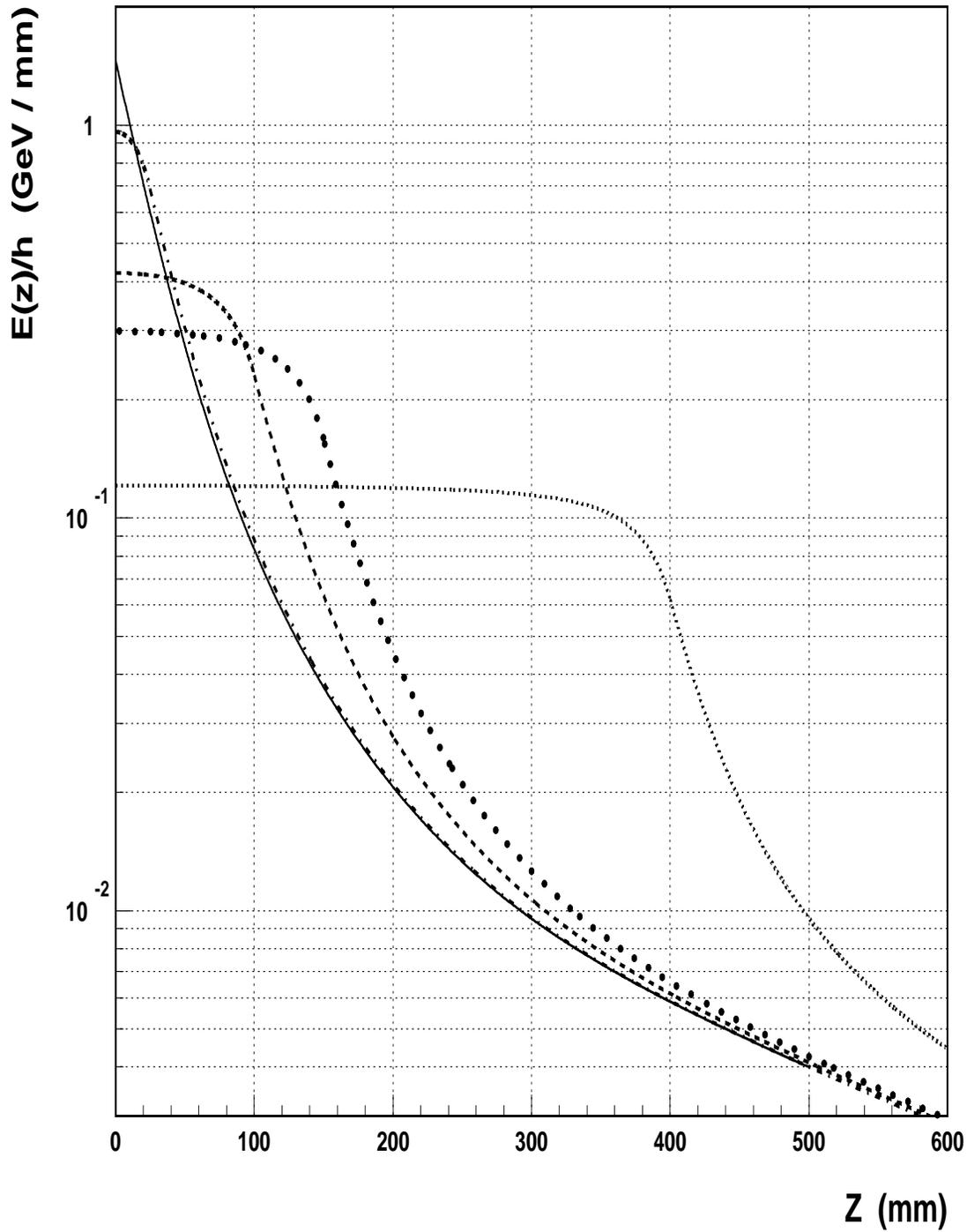,width=0.95\textwidth,height=0.85\textheight}}
        \\
        \end{tabular}
     \end{center}
      \caption{
        The calculated marginal density function $f(z)$ (the solid line)
        and the energy deposition function, $E(z)/h$, for various
        transverse sizes of a tower ($h$) :
        50 mm  (the dash-dotted line), 
        200 mm (the dashed line), 
        300 mm (the thick dotted line), 
        800 mm (the thin dotted line).
        The parameters for the entire calorimeter 
        (see Table \ref{Tb2})
        are used in the calculations.
       \label{fig:5-00}}
\end{figure*}

Figure \ref{fig:5-00} shows the calculated marginal density
function $f(z)$ and the energy deposition function, $E(z)/h$, at various
transverse sizes of tower $h = 50,\ 200,\ 300$ and 800 mm
using the obtained parameters for the entire calorimeter.
As a result of the volume integration,
the sharp $f(z)$ is transformed to the wide function $E(z)/h$, 
which clearly shows its relationship to the transverse width of a tower.
The values of FWHM are 40 mm for $f(z)$ and 204 mm for $E(z)/h$
at $h = 200$ mm.
Note that the transverse dimensions of a tower vary from 300 mm to 800 mm
for the different depth segments in final ATLAS Tile calorimeter design.
The difference between $f(0)$ and $E(0)/h$ becomes less then $5\%$ only
at $h$ less then 6 mm.

\begin{figure*}[tbph]
     \begin{center}
        \begin{tabular}{c}
\mbox{\epsfig{figure=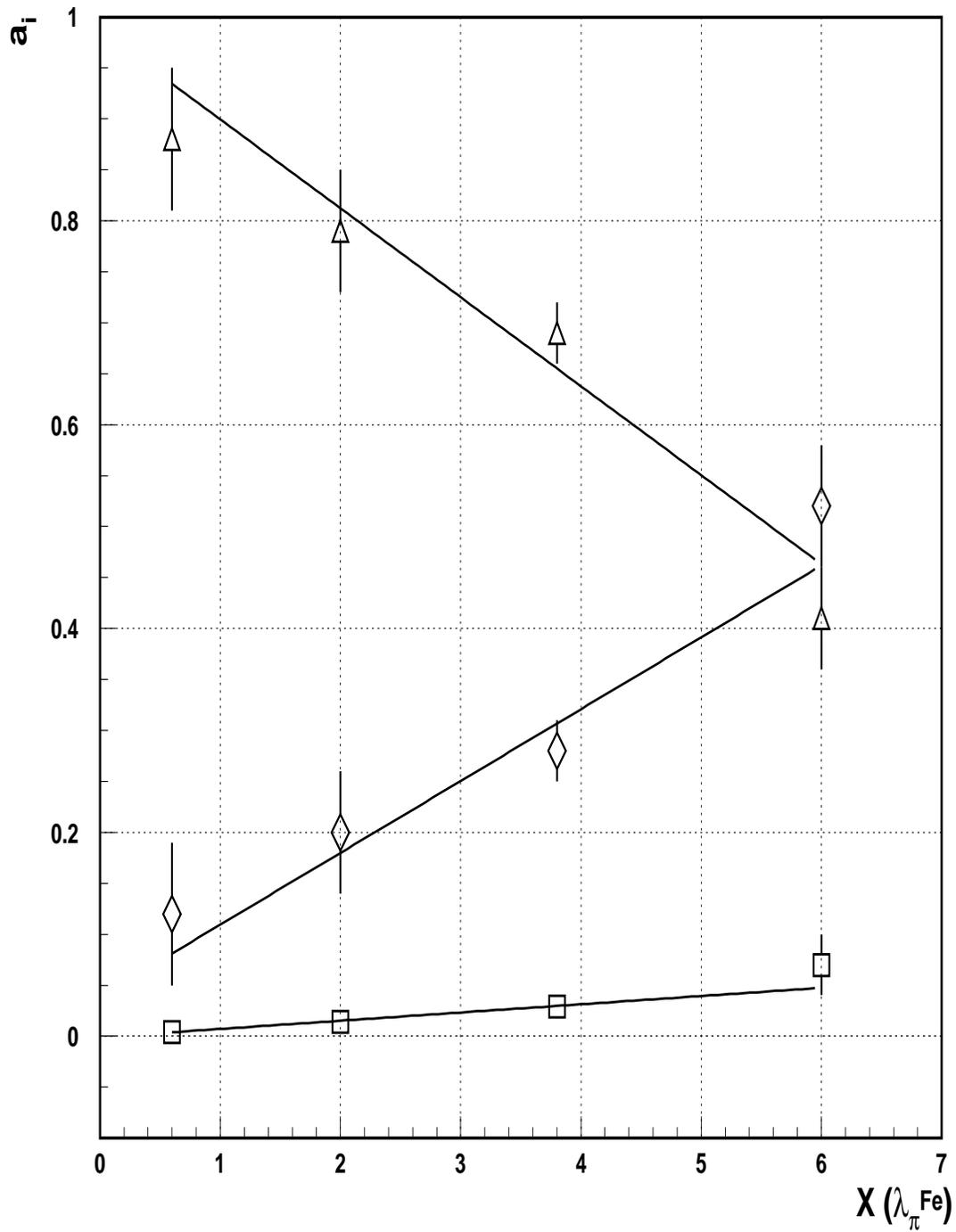,width=0.95\textwidth,height=0.85\textheight}}
        \\
        \end{tabular}
     \end{center}
      \caption{
        $X$ dependences of the parameters $a_i$:
        the triangles are the $a_1$ parameter,
        the diamonds are the $a_2$,
        the squares are the $a_3$.
       \label{fig:5}}
\end{figure*}

\begin{figure*}[tbph]
     \begin{center}
        \begin{tabular}{c}
\mbox{\epsfig{figure=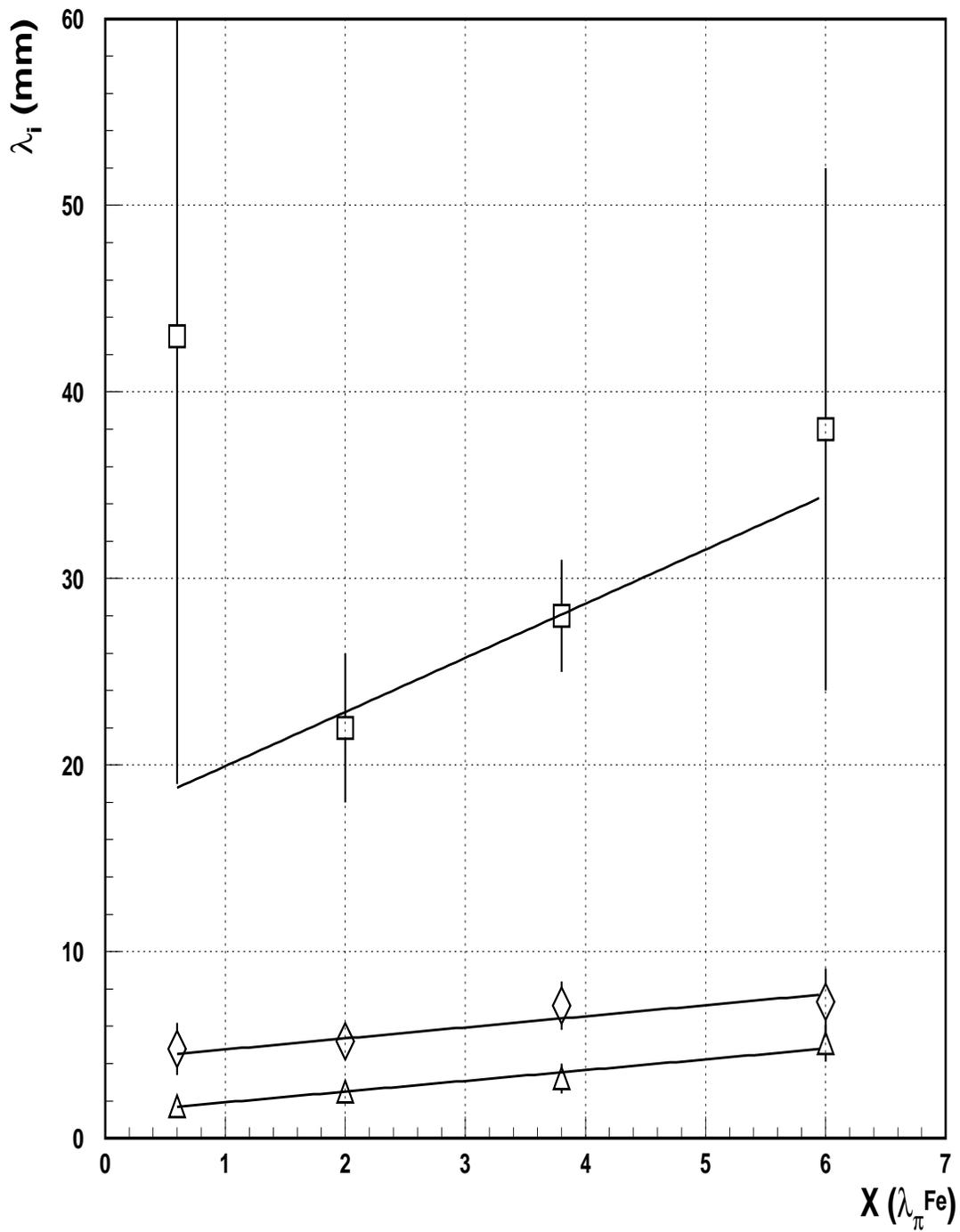,width=0.95\textwidth,height=0.85\textheight}}
        \\
        \end{tabular}
     \end{center}
      \caption{
        $X$ dependences of the parameters $\lambda_i$:
        the triangles are the $\lambda_1$ parameter,
        the diamonds are the $\lambda_2$,
        the squares are the $\lambda_3$.
       \label{fig:05}}
\end{figure*}

The parameters $a_i$ and $\lambda_i$
as a function of $x$ (in units of $\lambda_{\pi}^{Fe}$)
are displayed in Figs.\ \ref{fig:5} and \ref{fig:05}.
Here $\lambda_{\pi}^{Fe} = 207$ mm is the nuclear interaction length
for pions in iron 
\cite{budagov97}.
In these calculations, the effect of the $10^{\circ}$ incidence beam
angle has been corrected.
As can be seen from Figs.\ \ref{fig:5} and \ref{fig:05},
the value of $a_1$ decreases and the values of the remaining parameters,
$a_2, a_3$ and $\lambda_i$, increase as the shower develops.
This is a reflection of the fact that as the hadronic shower propagates 
into the calorimeter it becomes broader.
Note also that the $a_i$ and $\lambda_i$ parameters demonstrate 
linear behaviour as a function of $x$.
The lines shown are fits to the linear equations
\begin{equation}
        a_i (x) = \alpha_i + \beta_i  x 
\label{eai}
\end{equation}
and  
\begin{equation}
        \lambda_i (x) = \gamma_i + \delta_i  x.
\label{eli}
\end{equation}

The values of the parameters 
$\alpha_i$, $\beta_i$, $\gamma_i$ and $\delta_i$
are presented in Table \ref{Tb02}.
It is interesting to note that 
the linear behaviour of the slope exponential was also observed
for the low-density fine-grained flash chamber calorimeter
\cite{womersley88}.
However, some non-linear behaviour of the slope of the halo component 
was demonstrated for the uranium-scintillator $ZEUS$ calorimeter
at interaction lengths more than 5 $\lambda_I$
\cite{barreiro90}.

Similar results were obtained for the cumulative function distributions.
The cumulative function $F(z)$ is given by
\begin{equation}
\label{eq-pr-2}
F (z) = \sum_{k=1}^{4} F^{k}(z),
\end{equation}
where $F^{k}(z)$ is the cumulative function for depth segment $k$.
For each event, $F^{k}(z)$ is
\begin{equation}
\label{eq-pr-3}
F^{k} (z) =  \sum_{i=1}^{i_{max}}
                        \sum_{j=1}^{5} E_{ijk},
\end{equation}
where $i_{max} = 1, \ldots, 5$ is the last tower number in the sum.

\begin{figure*}[tbph]
     \begin{center}
        \begin{tabular}{cc}
\mbox{\epsfig{figure=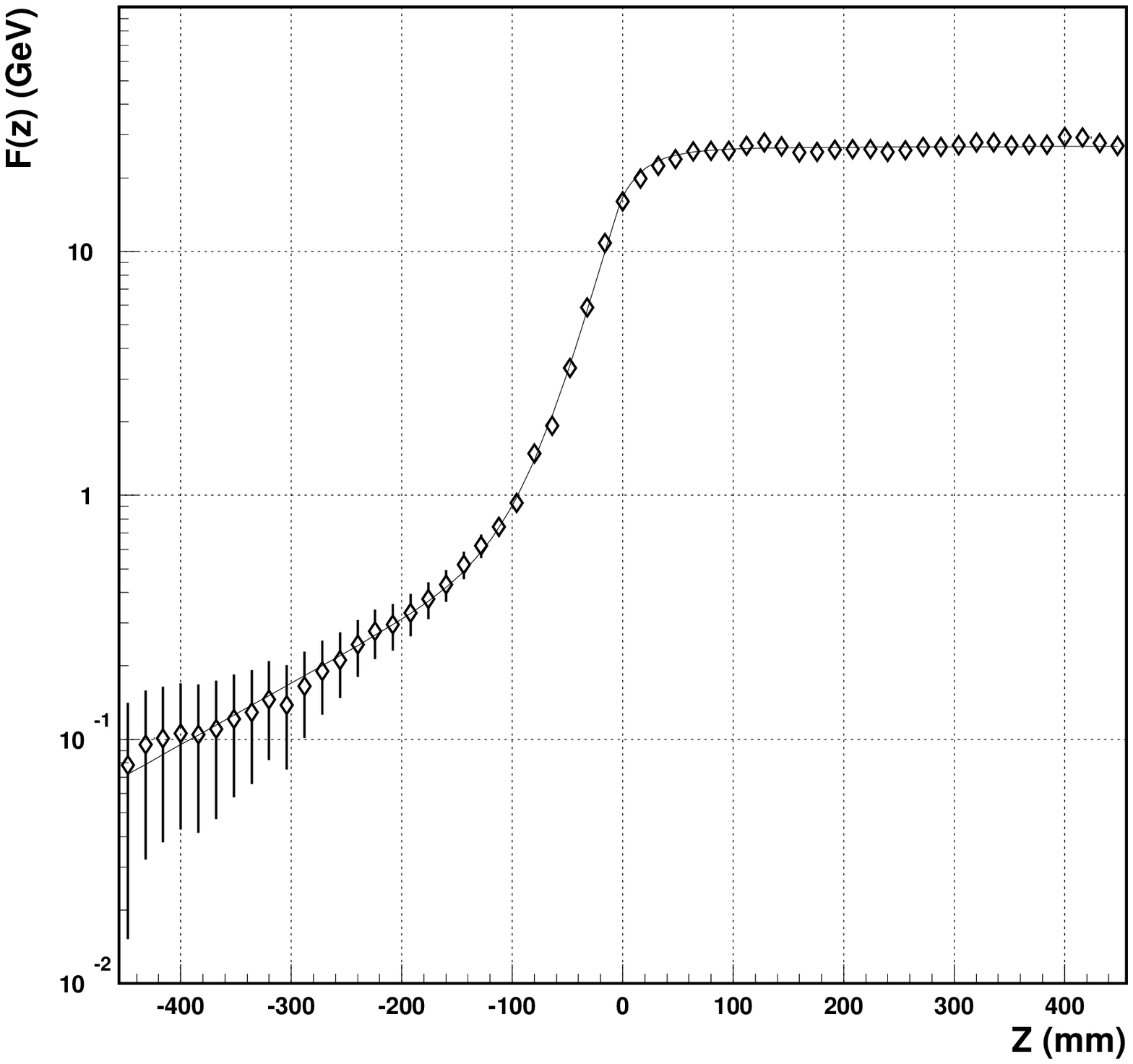,width=0.45\textwidth,height=0.4\textheight}}
&
\mbox{\epsfig{figure=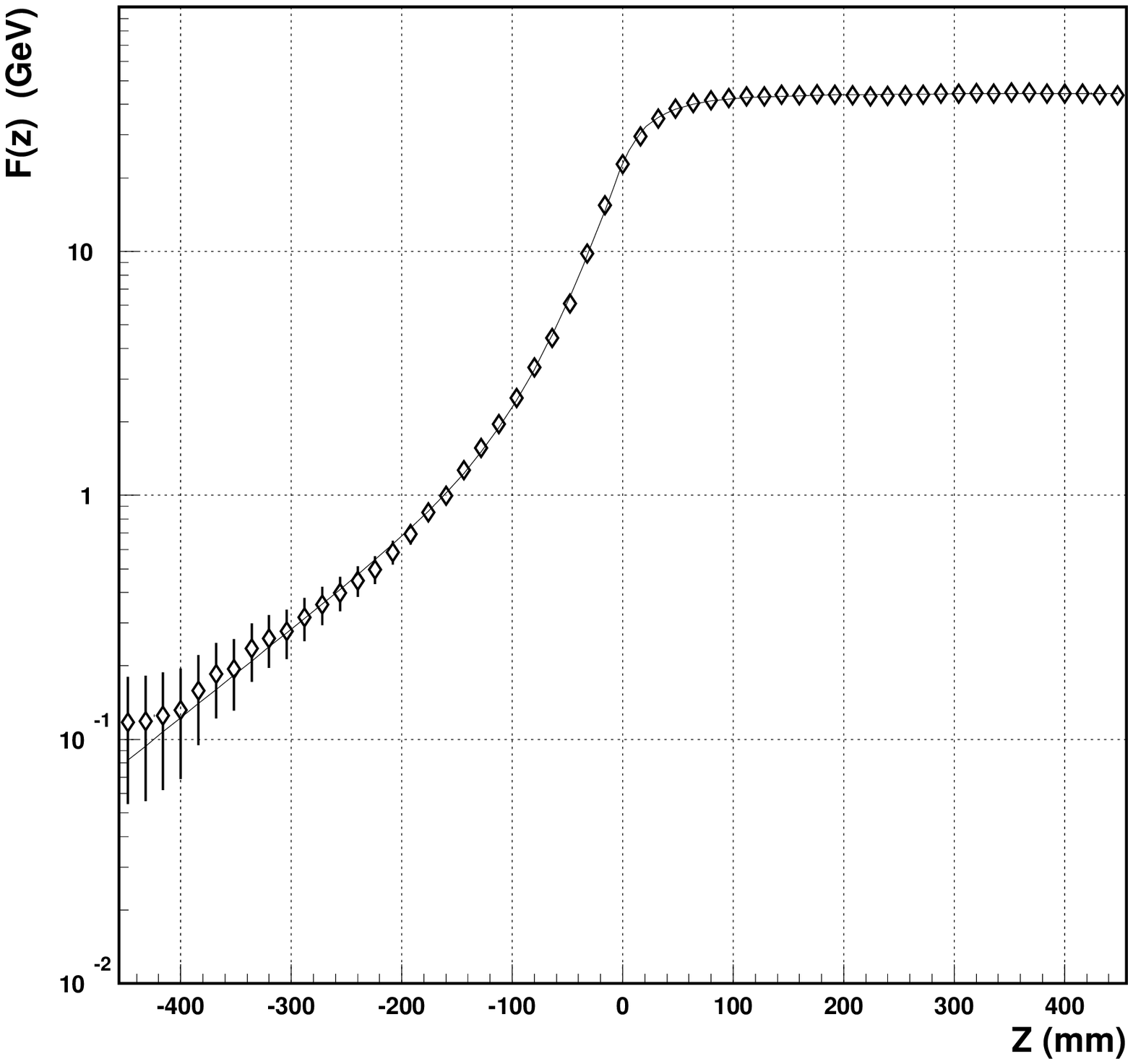,width=0.45\textwidth,height=0.4\textheight}}
        \\
\mbox{\epsfig{figure=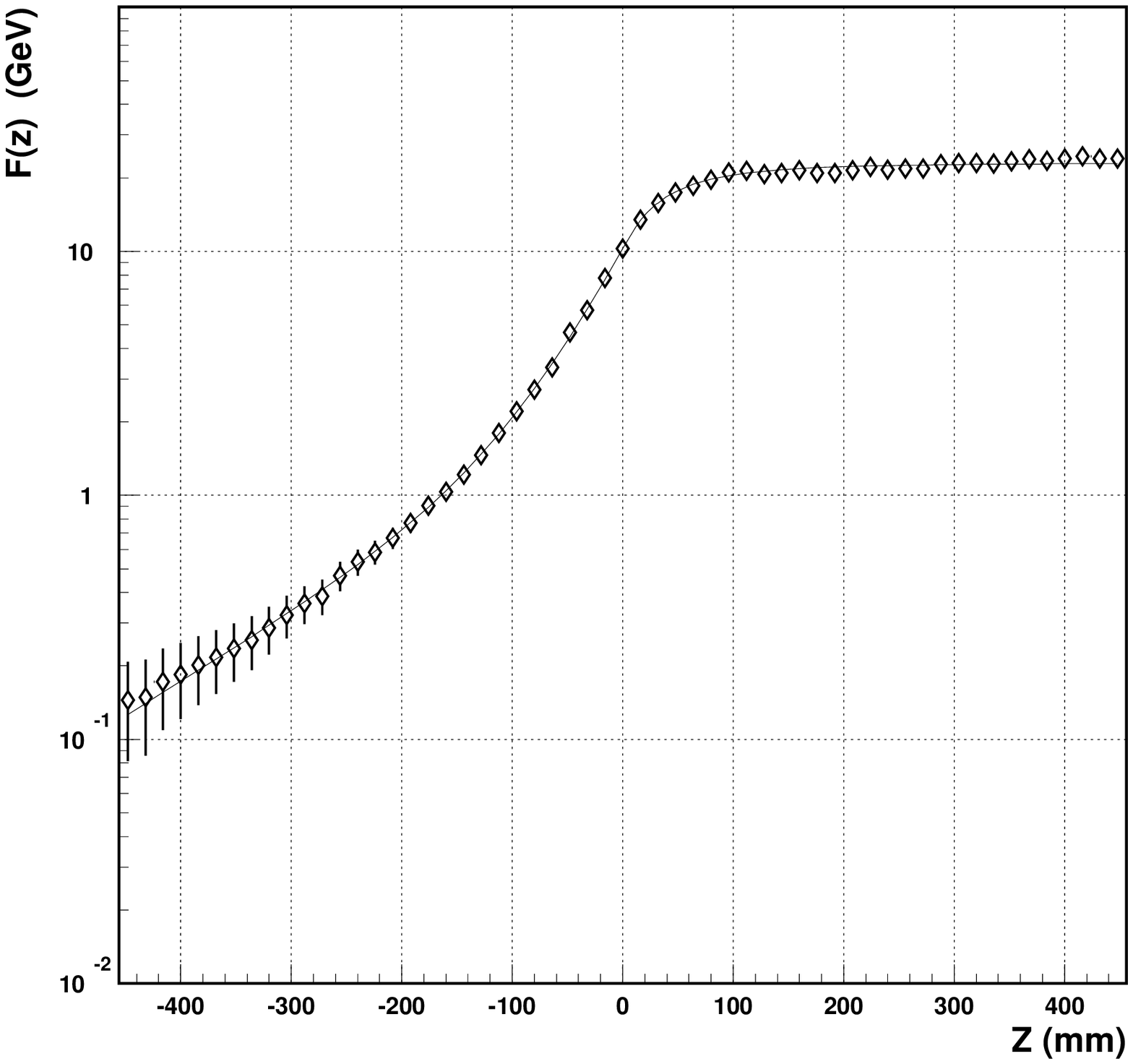,width=0.45\textwidth,height=0.4\textheight}}
&
\mbox{\epsfig{figure=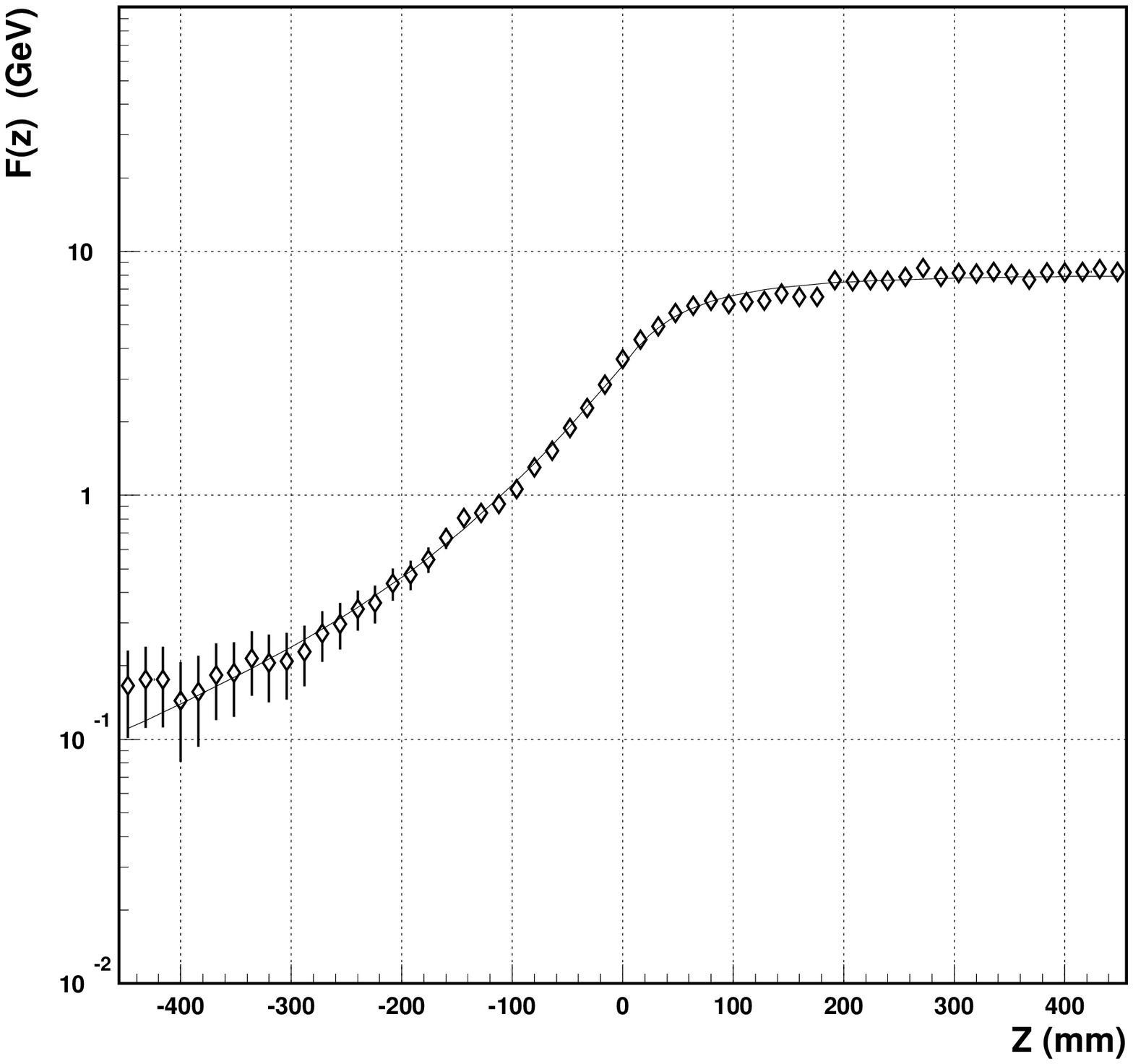,width=0.45\textwidth,height=0.4\textheight}}
        \\
        \end{tabular}
     \end{center}
      \caption{
        Cumulative functions $F(z)$ for depth segments 1 -- 4 
        as a function of the $z$ coordinate:
        top left is for depth segment 1,
        top right is for depth segment 2,
        bottom left is for depth segment 3,
        bottom right is for depth segment 4. 
        Statistical and systematic errors, summed in quadrature, are shown. 
        Curves are fits of equations (16) and (\ref{e23-2}) to the data.
       \label{fig:f7-001}}
\end{figure*}

\begin{figure*}[tbph]
     \begin{center}
        \begin{tabular}{c}
\mbox{\epsfig{figure=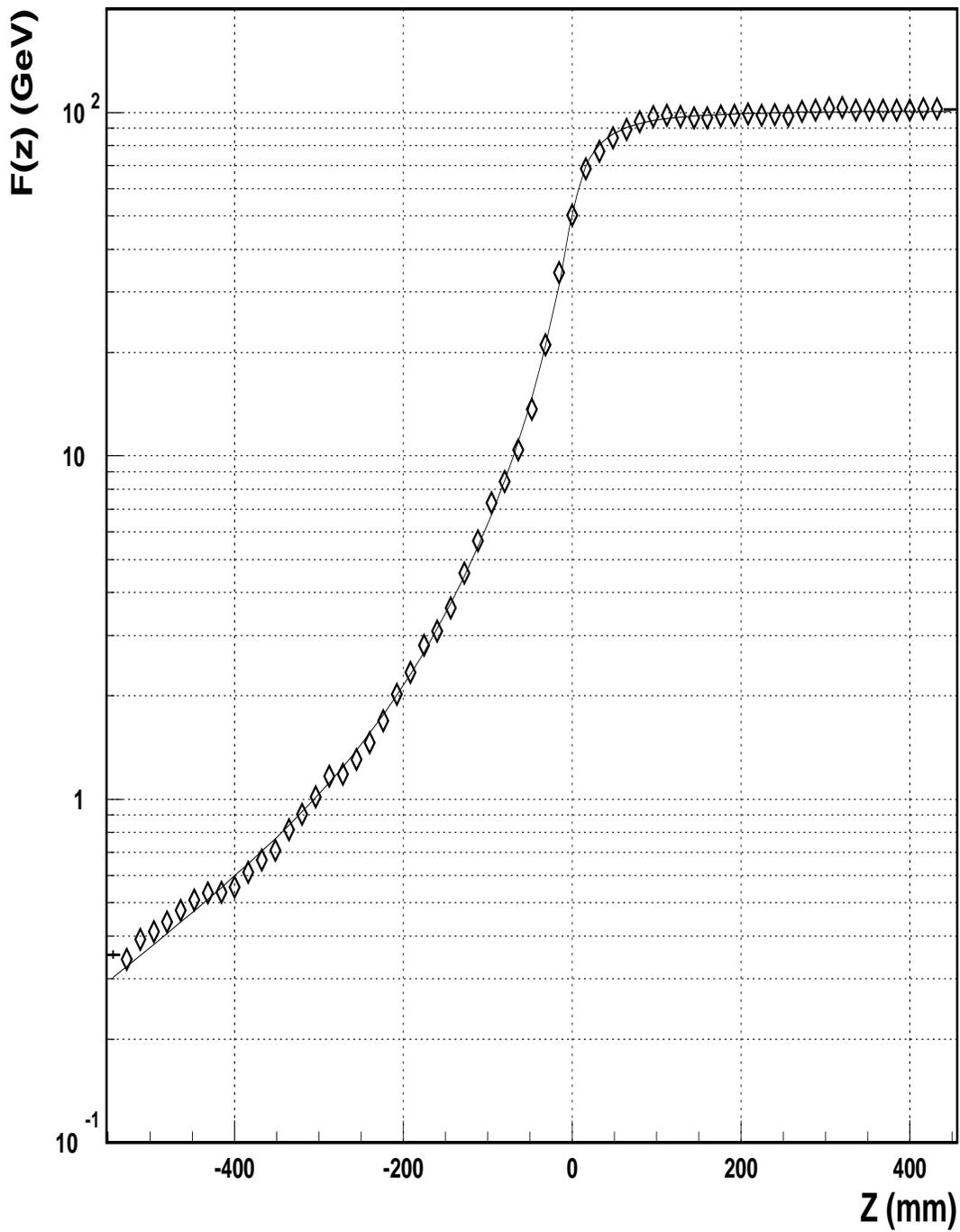,width=0.95\textwidth,height=0.85\textheight}}
        \\
        \end{tabular}
     \end{center}
      \caption{
        The cumulative function $F(z)$ for the entire calorimeter
        as a function of the $z$ coordinate.
        Only statistical errors are shown.
        Curves are fits of equations (16) and (\ref{e23-2}) to the data.
       \label{fig:f9-1}}
\end{figure*}

Figures \ref{fig:f7-001} and \ref{fig:f9-1} present the cumulative 
functions for four depth segments and for the entire calorimeter.
The curves are fits of equations (16) and (\ref{e23-2}) to the data.
Systematic and statistical errors are again added in quadrature. 
The results of the cumulative function fits are less reliable 
and in what follows we use the results from energy depositions in a tower.

However, the marginal density functions determined by the two methods
(by using the energy deposition spectrum and the cumulative function)
are in reasonable agreement.

\section{Radial Hadronic Shower Energy Density}

\begin{figure*}[tbph]
     \begin{center}
        \begin{tabular}{cc}
\mbox{\epsfig{figure=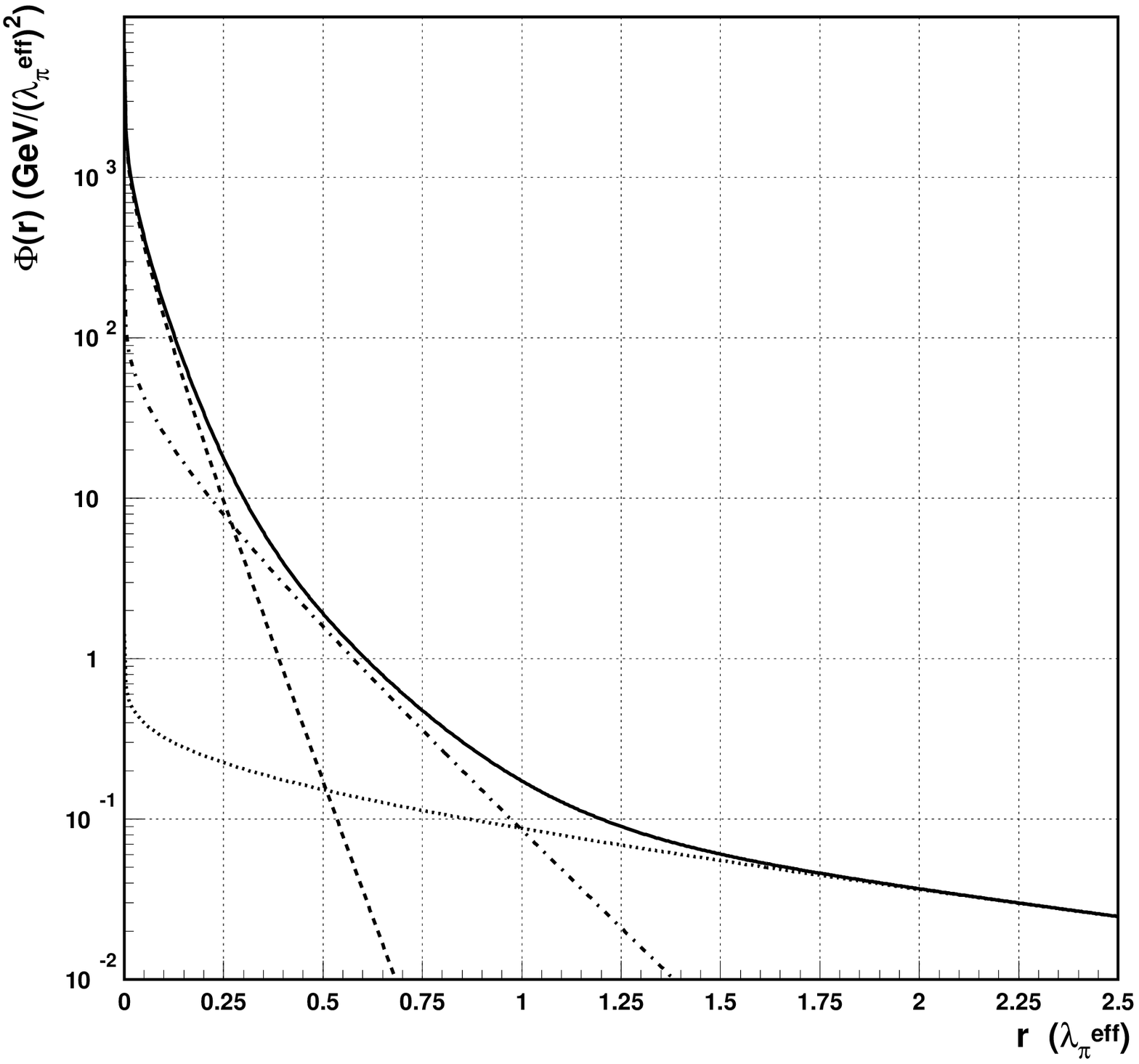,width=0.45\textwidth,height=0.4\textheight}}
&
\mbox{\epsfig{figure=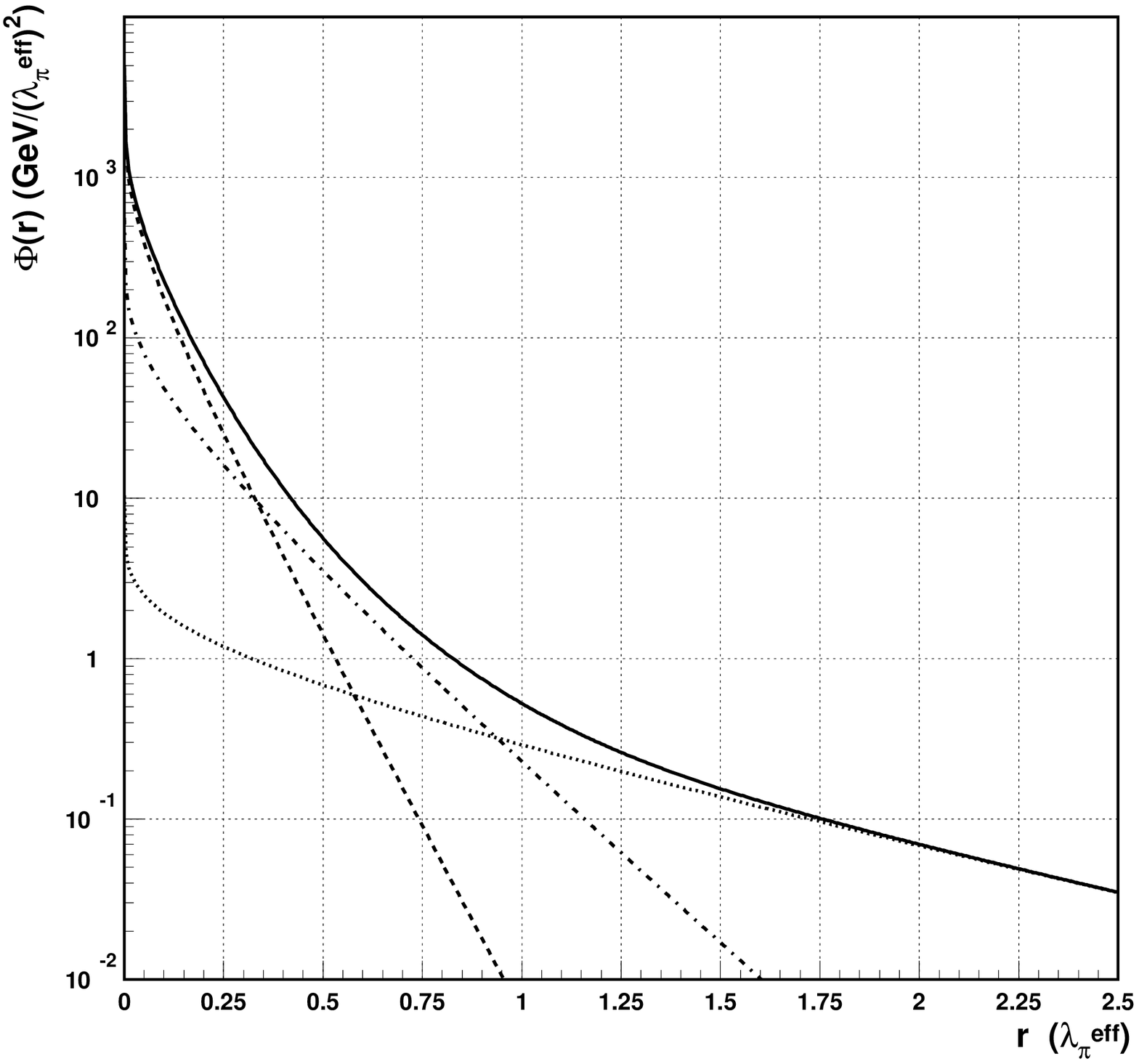,width=0.45\textwidth,height=0.4\textheight}}
        \\
\mbox{\epsfig{figure=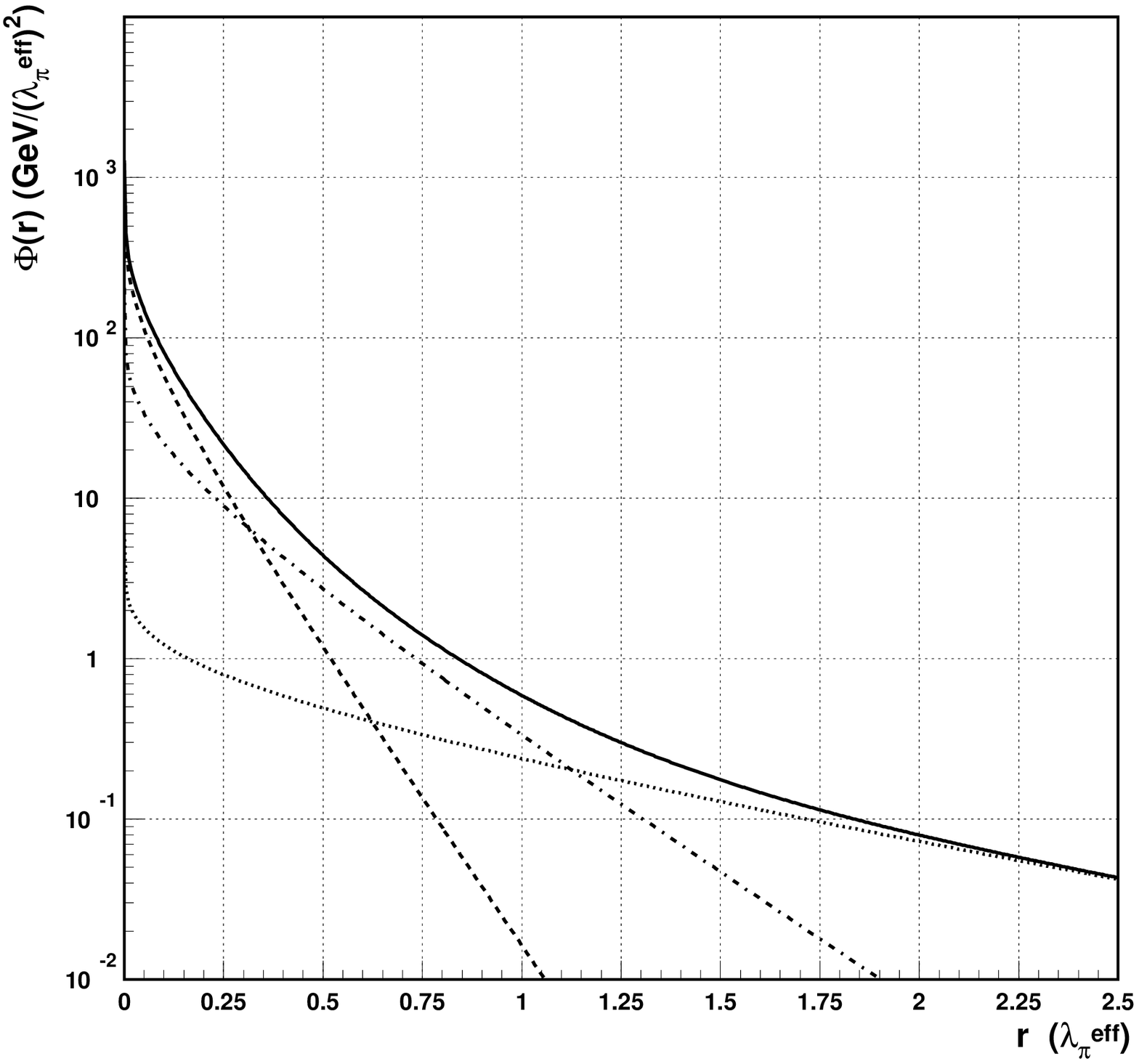,width=0.45\textwidth,height=0.4\textheight}}
&
\mbox{\epsfig{figure=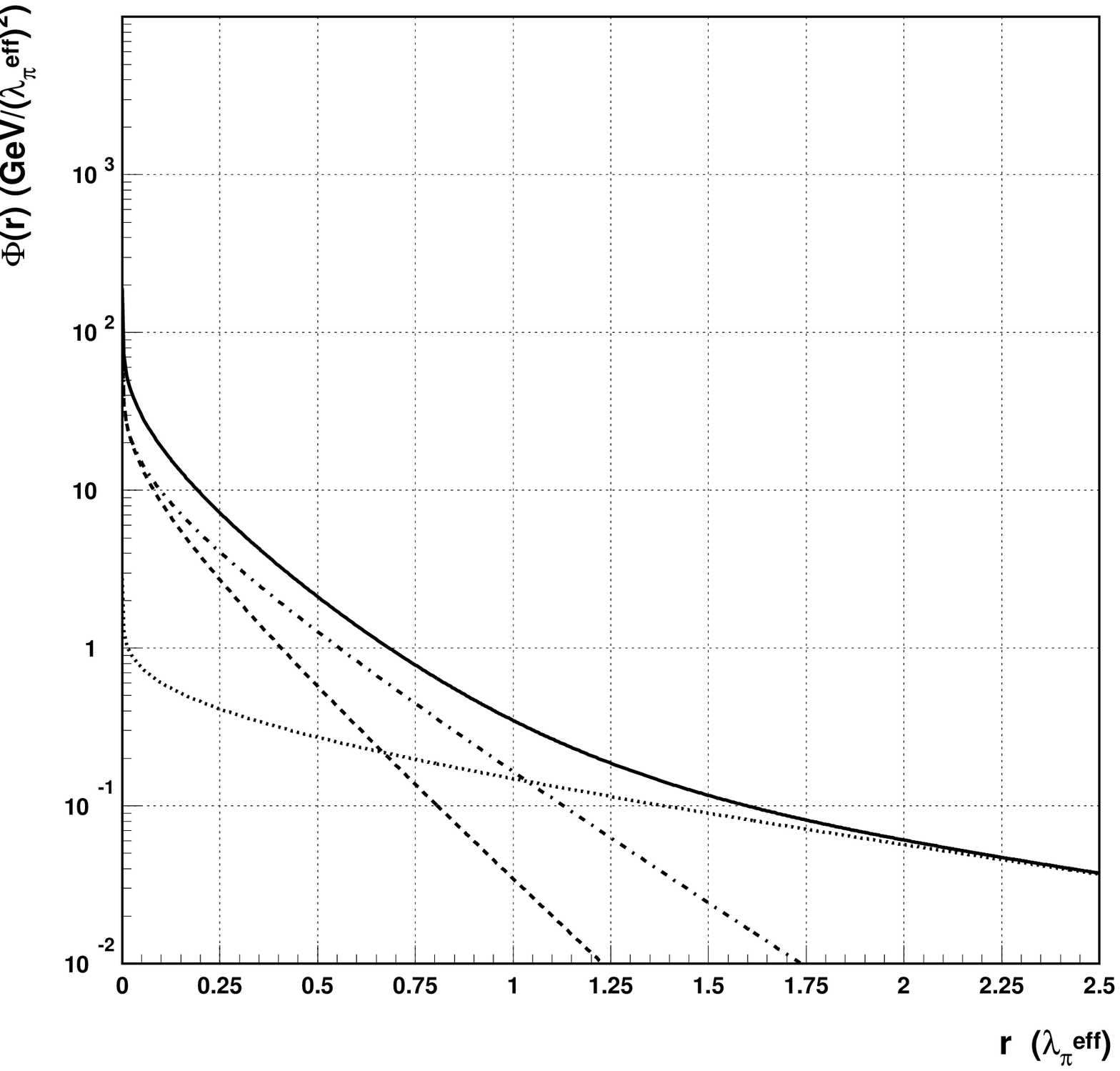,width=0.45\textwidth,height=0.4\textheight}}
        \\
        \end{tabular}
     \end{center}
      \caption{
        Radial energy density, $\Phi(r)$, as a function of $r$ for Tile 
        calorimeter
        for depth segments 1 -- 4:
        top left is for depth segment 1,
        top right is for depth segment 2,
        bottom left is for depth segment 3,
        bottom right is for depth segment 4. 
        The solid lines are the energy densities $\Phi(r)$,
        the dashed lines are the contribution from the first term from 
        equation (\ref{e23}),
        the dash-dotted lines are the contribution from the second term,
        the dotted lines are the contribution from the third term.
       \label{fig:f17-001}}
\end{figure*}

\begin{figure*}[tbph]
     \begin{center}
        \begin{tabular}{c}
\mbox{\epsfig{figure=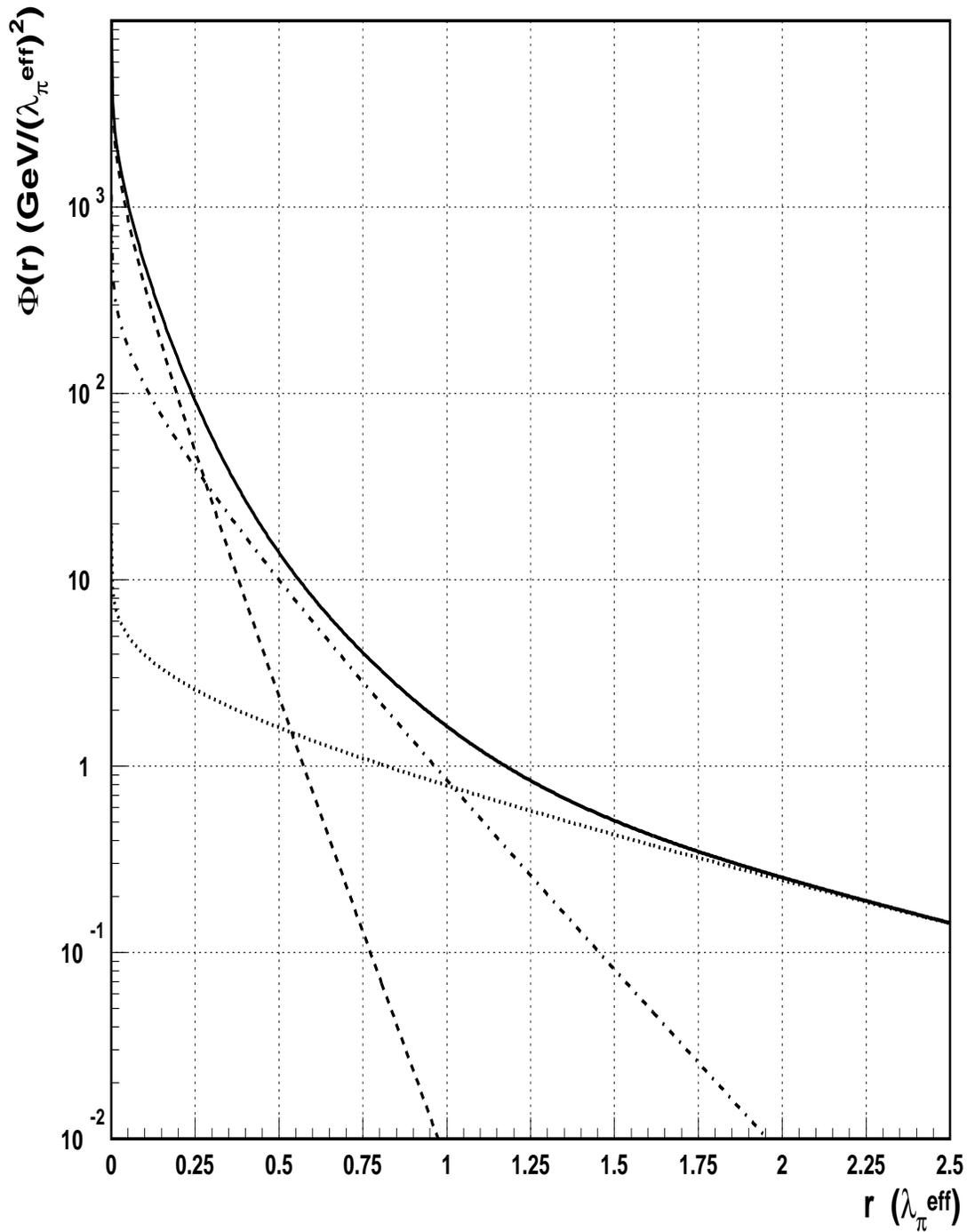,width=0.95\textwidth,height=0.85\textheight}}
        \\
        \end{tabular}
     \end{center}
       \caption{
        The radial energy density as a function of $r$ 
        (in units of $\lambda_{\pi}^{eff}$) for Tile calorimeter 
        (the solid line),
        the contribution to $\Phi(r)$ 
        from the first term in equation (\ref{e23}) (the dashed line),
        the contribution to $\Phi(r)$ 
        from the second term (the dash-dotted line),
        the contribution to $\Phi(r)$ 
        from the third term (the dotted line).
       \label{fig:f10-1}}
\end{figure*}

Using formula (\ref{e23}) and the values of the parameters 
$a_i$, $\lambda_i$, given in Table~\ref{Tb2}, 
we have determined the underlying radial 
hadronic shower energy density functions, $\Phi(r)$.
The results are shown in Figure \ref{fig:f17-001} for depth segments 1 -- 4 
and in Figure \ref{fig:f10-1} for the entire calorimeter.
The contributions of the three terms of $\Phi(r)$ are also shown.

The functions $\Phi(r)$ for separate depth segments of a calorimeter are given
in this paper for the first time. The function $\Phi(r)$ for the entire
calorimeter was previously given for a Lead-scintillating fiber calorimeter
\cite{acosta92}; it is given here for the first time for an Iron-scintillator
calorimeter. The analytical functions giving the radial energy density 
for different depth segments allow to easily obtain 
the shower energy deposition in any calorimeter cell, 
shower containment fractions and the cylinder radii for any 
given shower containment fraction.

\begin{figure*}[tbph]
     \begin{center}
        \begin{tabular}{c}
\mbox{\epsfig{figure=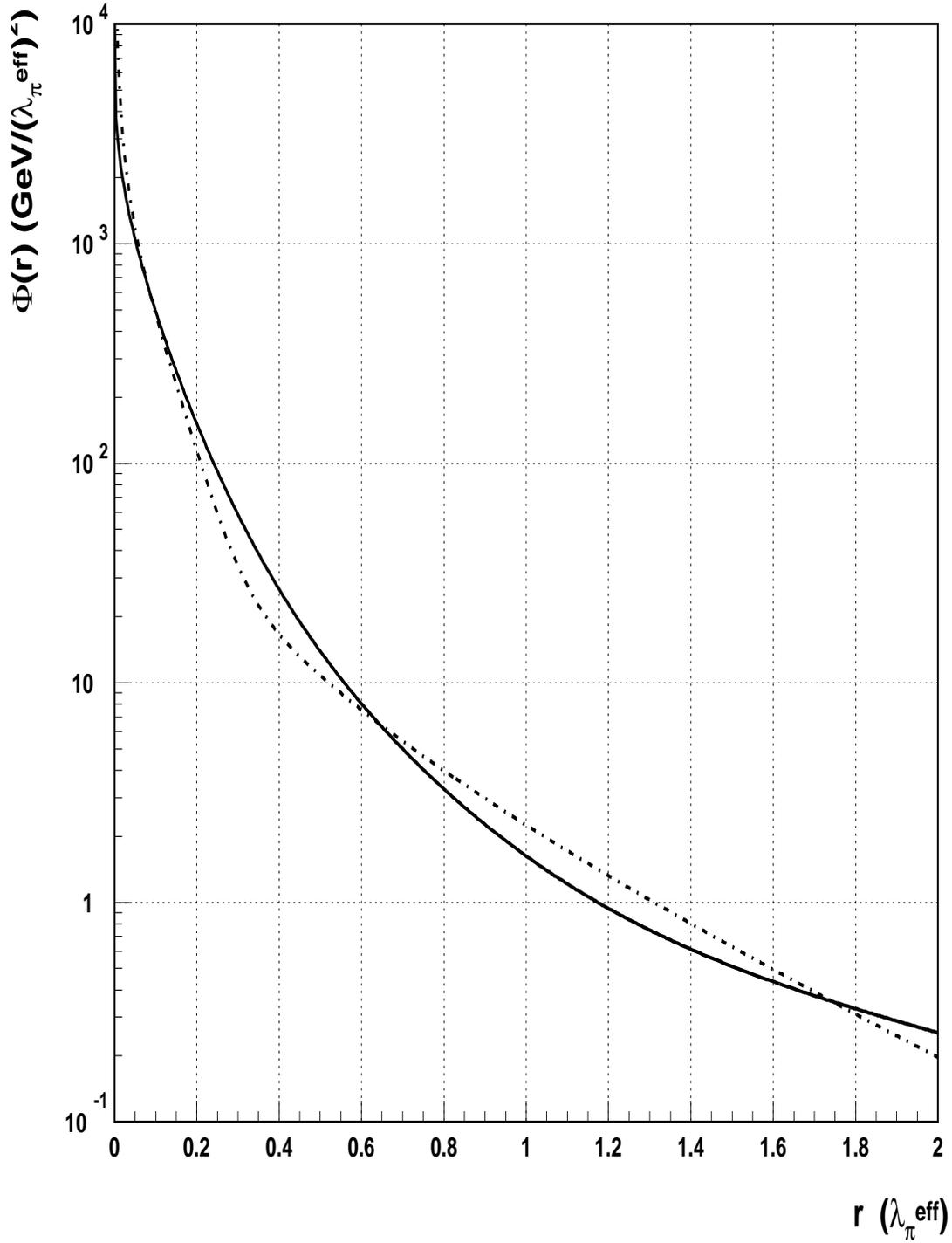,width=0.95\textwidth,height=0.85\textheight}}
        \\
        \end{tabular}
     \end{center}
       \caption{
        Comparison of the radial energy densities
        as a function of $r$ (in units of $\lambda_{\pi}^{eff}$) for
        Tile calorimeter (the solid line) and 
        lead-scintillating fiber calorimeter 
        (the dash-dotted line).
       \label{fig:f10-2}}
\end{figure*}

The function $\Phi(r)$ for the entire calorimeter has been compared 
with the one for the lead-scintillating fiber calorimeter 
of ref.\ \cite{acosta92}, that has
about the same effective nuclear interaction length for pions 
(namely 251 mm for the tile and 244 mm for the fiber 
calorimeter \cite{budagov97}).
The two
radial density functions are rather similar as seen 
in Fig.\ \ref{fig:f10-2}.
The lead-scintillating fiber calorimeter density function 
$\Phi(r)$, which 
was obtained from 
a 80 GeV $\pi^{-}$ grid scan at an angle 
of $2^{\circ}$ with respect to the fiber direction,
was parametrized 
using formula (\ref{e1}) with
$b_1 = 0.169$ pC/mm, $b_2 = 0.677$ pC/mm, ${\mu}_1 = 140$ mm and
${\mu}_2 = 42.4$ mm.
For the sake of comparing the radial density functions of the two calorimeters,
the distribution from \cite{acosta92} was normalised to the $\Phi(r)$ of the
Tile calorimeter.
Precise agreement between these functions
should not be expected because of 
the effect of the 
different absorber materials
used in the two detectors
(e.\  g.\  the radiation/interaction length ratio for the Tile calorimeter 
is three times larger than for lead-scintillating fiber calorimeter 
\cite{budagov97}),
the values of $e/h$ are different, as is hadronic activity of showers because
fewer neutrons are produced 
in iron than in lead \cite{Fabjan82}, \cite{gabriel94}). 

\section{Radial Containment}

\begin{figure*}[tbph]
     \begin{center}
        \begin{tabular}{c}
\mbox{\epsfig{figure=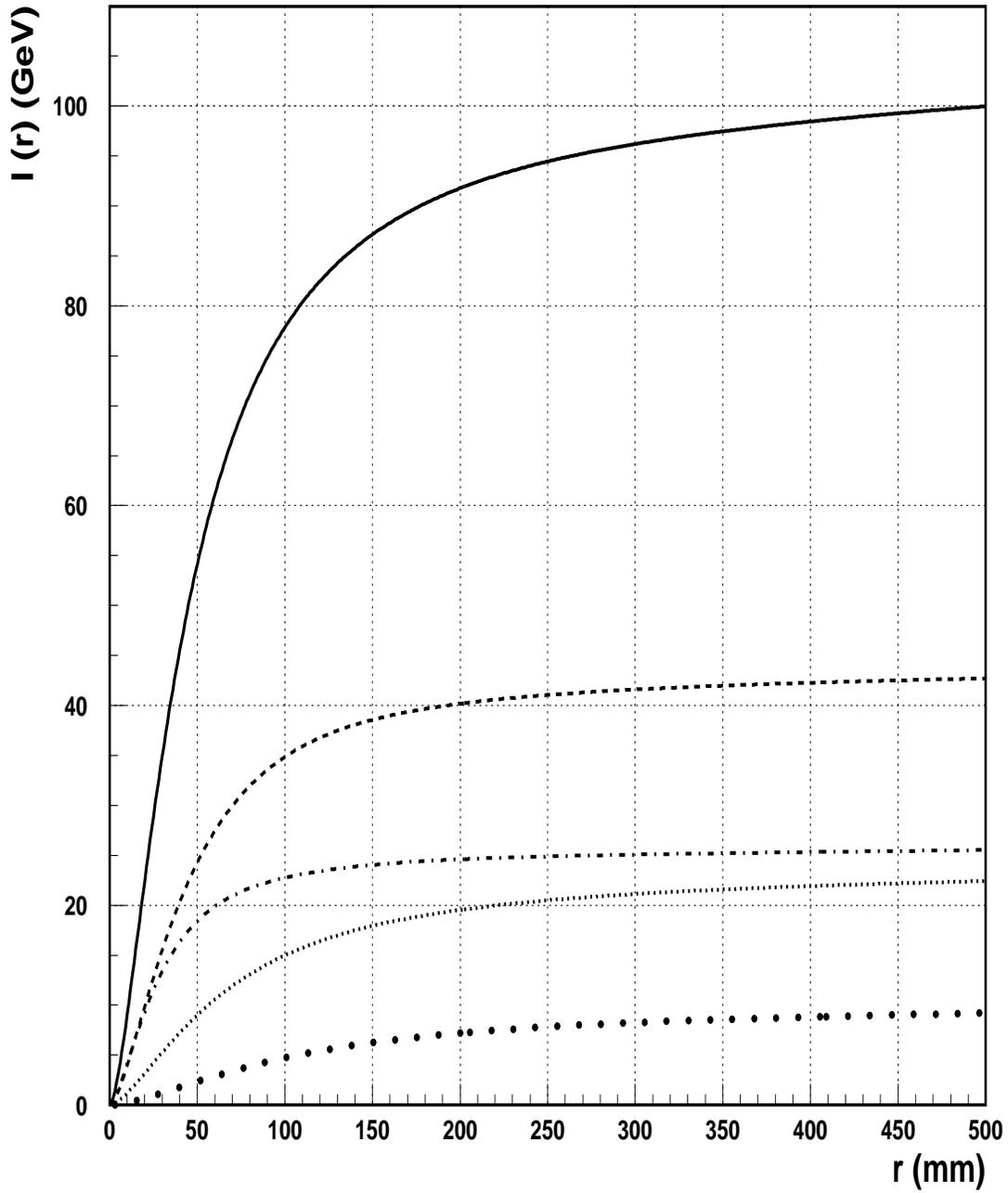,width=0.95\textwidth,height=0.80\textheight}}
        \\
        \end{tabular}
     \end{center}
       \caption{
        Containment of shower $I(r)$ (the solid line)
        as a function of radius for the entire Tile calorimeter.
        The dash-dotted line is the contribution from the first depth segment,
        the dashed line is the contribution from the second depth segment,
        the thin dotted line is the contribution from the third depth segment,
        the thick dotted line is the contribution from the fourth 
        depth segment.
       \label{fig:f11-1}}
\end{figure*}

An other issue on which new results are presented here 
is the longitudinal development of 
shower transverse dimensions.
The parametrization of the radial density function, $\Phi(r)$,
was integrated to yield the shower containment as a function 
of the radius, $I(r)$.
Figure \ref{fig:f11-1} shows the transverse containment of the pion shower,
$I(r)$, as a function of $r$ for four depth segments 
and for the entire calorimeter.

\begin{figure*}[tbph]
     \begin{center}
        \begin{tabular}{c}
\mbox{\epsfig{figure=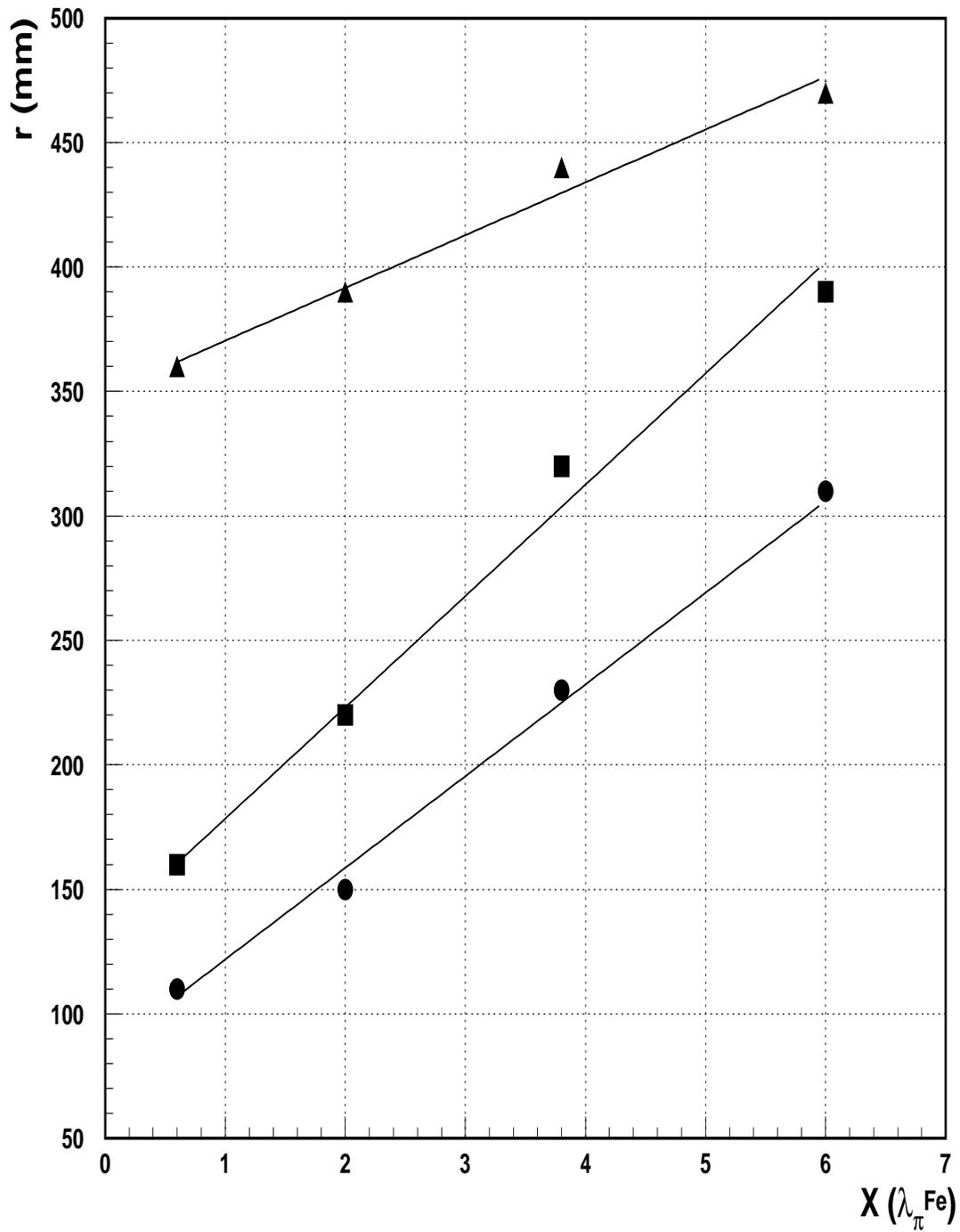,width=0.95\textwidth,height=0.85\textheight}}
        \\
        \end{tabular}
     \end{center}
       \caption{
        The radii of cylinders for the given shower containment
        as a function of depths: 
        the black circles are $90\%$ of containment, 
        the black squares are $95\%$,
        the black triangles are $99\%$.
       \label{fig:f11-2}}
\end{figure*}

In Table \ref{Tb3} and Fig.\ \ref{fig:f11-2} 
the radii of cylinders for the given shower containment
($90\%$, $95\%$ and $99\%$) extracted from Fig.\ \ref{fig:f11-1}
as a function of depth are shown.
The centers of depth segments, 
$x$, are given in units of $\lambda_{\pi}^{Fe}$.
Solid lines are the linear fits to the data:
$r(90\%) = (85 \pm 6) + (37 \pm 3)x$,
$r(95\%) = (134 \pm 9) + (45 \pm 3)x$, 
$r(99\%) = (349 \pm 7) + (22 \pm 2)x$ (mm).
As can be seen, these containment  radii increase linearly with depth.
Such a linear increase of $95\%$ lateral shower containment 
with depth is also observed 
in an other iron-scintillator calorimeter at 50 and 140 GeV 
\cite{holder78}.
It is interesting to note that the shower radius for $95\%$ radial 
containment for the entire calorimeter is equal to 
$\lambda_{\pi}^{eff} = 251$ mm \cite{budagov97}
which justifies the frequently encountered statement that 
$r(95\%) \approx \lambda_I$ \cite{Fabjan82},
where $\lambda_I$ is $\lambda_{\pi}^{eff}$ in our case.
For the entire Tile calorimeter the $99\%$ containment radius is equal to 
$1.7 \pm 0.1$ $\lambda_{\pi}^{eff}$.

Based on our study, we believe that it is a poor approximation to regard
the values obtained from the marginal density function  
or the energy depositions in strips as the measure of the transverse 
shower containment, as was done in 
\cite{womersley88}.
In that paper the value of $1.1\ \lambda_{\pi}^{eff}$
was obtained for $99 \%$ containment at 100 GeV, 
and the conclusion was drawn that 
their ``result is consistent with the {\it rule of thumb} 
that a shower is contained within a cylinder of radius equal 
to the interaction length of a calorimeter material''.
However Tile calorimeter measurements show that the cylinder radius
for $99\%$ shower containment is about two interaction lengths.
If we extract the lateral shower containment dimension
using instead the integrated function 
$F(z)$, given in Fig.\ \ref{fig:f9-1}, 
we obtain the value of $300$ mm or $1.2\ \lambda_{\pi}^{eff}$,
which agrees with \cite{womersley88}.

\section{Longitudinal Profile}

We have examined the differential deposition of energy $\Delta E/ \Delta x$ 
as a function of $x$, the distance along the shower axis.
Table \ref{long-1} lists
the centers in $x$ of the depth segments, $x$, 
and the lengths along $x$ of the depth segments, 
$\Delta x$, in units of $\lambda_{\pi}^{Fe}$,
the average shower energy depositions in various depth segments, $E_{0}$, 
and the energy depositions 
per interaction length 
$\lambda_{\pi}^{Fe}$, $\Delta E / \Delta x$. 
Note that the values of $E_0$ have been obtained taking into account
the longitudinal energy leakage which amounts to 1.8 GeV for 100 GeV
\cite{budagov96-76}.

\begin{figure*}[tbph]
     \begin{center}
        \begin{tabular}{c}
\mbox{\epsfig{figure=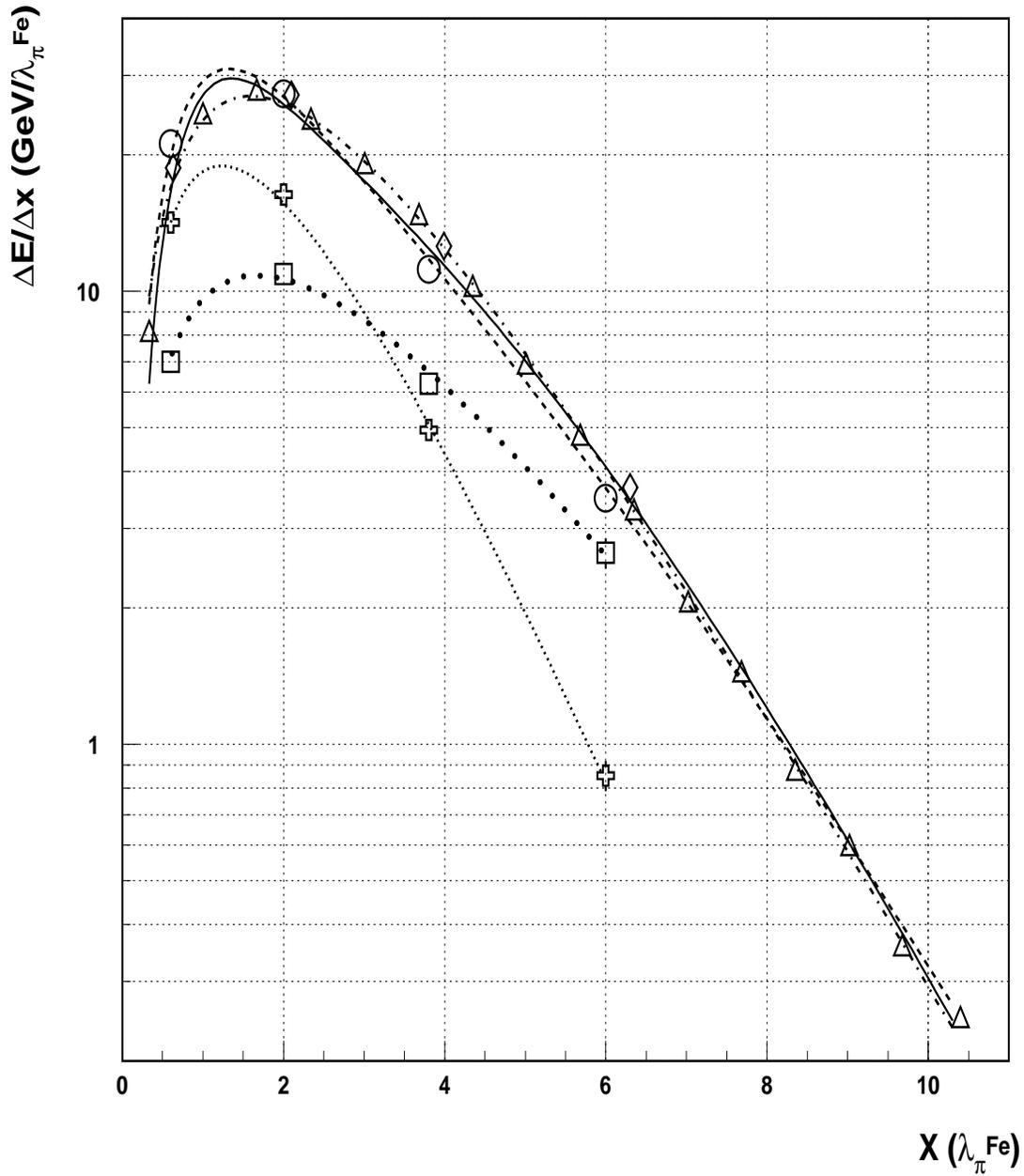,width=0.95\textwidth,height=0.75\textheight}} 
        \\
        \end{tabular}
     \end{center}
       \caption{
        The longitudinal profile (circles)
        of the hadronic shower at 100 GeV as a function of the
        longitudinal coordinate $x$ in units of $\lambda_{\pi}^{Fe}$.
        Open triangles are data from the calorimeter of 
        Ref.\ \cite{hughes90}, 
        diamonds are the Monte Carlo (GEANT-FLUKA) predictions. 
        The dash-dotted line is the fit by function (\ref{elong00}),
        the solid line is calculated with function (\ref{elong2}) 
        with parameters from Ref.\ \cite{hughes90},
        the dashed line is calculated with function (\ref{elong2}) 
        with parameters from Ref.\ \cite{bock81}. 
        The electromagnetic and hadronic components of the shower
        (crosses and squares), together with their fits using
        (\ref{elong00}), are discussed in Section 9.
       \label{fig:f15}}
\end{figure*}

Our values of  $\Delta E / \Delta x$
together with the data of \cite{hughes90}
and Monte Carlo predictions (GEANT-FLUKA + MICAP) \cite{juste95} are
shown
in Fig.\ \ref{fig:f15}.
The longitudinal energy deposition
for our calorimeter using longitudinal orientation of the scintillating
tiles is in good agreement with that of a 
conventional iron-scintillator
calorimeter.

The longitudinal profile, $\Delta E / \Delta x$, may be approximated
using two parametrizations.
The first form is 
\begin{equation}
        \frac{dE(x)}{dx} = 
                \frac{E_{f}\ \beta^{\alpha + 1}}{\Gamma (\alpha + 1)}\
                x^{\alpha}\ e^{-\beta x}
\label{elong00}
\end{equation}
where $E_{f} = E_{beam}$, and $\alpha$ and $\beta$ are free parameters.
Our data at 100 GeV and those of Ref.\ \cite{hughes90} at 100 GeV were
jointly fit to this expression; the fit is shown in Fig.\ \ref{fig:f15}.

The second form is the analytical representation of the
longitudinal shower profile from the front of the calorimeter
\begin{eqnarray}
        \frac{dE (x)}{dx} & = & 
                N\ 
                \Biggl\{
                \frac{w\ X_{0}}{a}\ 
                \biggl( \frac{x}{X_{0}} \biggr)^a\ 
                e^{- b \frac{x}{X_{0}}}\
                {}_1F_1 \biggl( 1,\ a+1,\ 
                \biggl(b - \frac{X_{0}}{\lambda_I} \biggr)\ \frac{x}{X_{0}}
                \biggr) 
                \nonumber \\
                & & + \ 
                \frac{(1 - w)\ \lambda_I}{a}\ 
                \biggl( \frac{x}{\lambda_I} \biggr)^a\ 
                e^{- d \frac{x}{\lambda_I}}\
                {}_1F_1 \biggl( 1,\ a+1,\
                \bigl( d - 1 \bigr)\ \frac{x}{\lambda_I} \biggr)
                \Biggr\} , 
\label{elong2}
\end{eqnarray}
where ${}_1F_1$ is the confluent hypergeometric function
\cite{abramovitz64}. Here the depth variable, $x$, is the depth in
equivalent $Fe$, $X_0$ is the radiation length in $Fe$
and in this case $\lambda_I$ is $\lambda_\pi^{Fe}$.
The normalisation factor, $N$, is given by
\begin{equation}
        N =     
                E_{beam} \biggl/ \int \limits_{0}^{\infty} \frac{dE(x)}{dx} dx
          =     \frac{E_{beam}}{\lambda_I\ \Gamma (a)\ 
                \bigl(w\ X_{0}\ b^{-a} + (1-w)\ \lambda_I\ d^{-a} \bigr)
                } . 
\label{elong02}
\end{equation}
This form was suggested in \cite{kulchitsky98} and derived by integration  
over the shower vertex positions of the longitudinal shower development
from the shower origin
\begin{equation}
        \frac{dE (x)}{dx} = 
        \int \limits_{0}^{x} 
        \frac{dE_{s}(x-x_v)}{dx}\ e^{- \frac{x_v}{\lambda_I}}\ dx_v ,
\label{elong1}
\end{equation}
where $x_v$ is the coordinate of the shower vertex.
(This is necessary because with the Tile calorimeter
longitudinal segmentation the shower vertex is not measured).
For the para\-me\-trization of longitudinal shower development,
the well known parametrization 
suggested by Bock et al.\ \cite{bock81} has been used
\begin{equation}
        \frac{dE_{s} (x)}{dx} = 
                N\      
                \Biggl\{
                w\ \biggl( \frac{x}{X_{0}} \biggr)^{a-1}\ 
                e^{- b \frac{x}{X_{0}}}\
                + \ 
                (1-w)\ 
                \biggl( \frac{x}{\lambda_I} \biggr)^{a-1}\ 
                e^{- d \frac{x}{\lambda_I}}
                \Biggr\}, 
\label{elong0}
\end{equation}
where $a,\ b,\ d,\ w$ are parameters.

We compare the form (\ref{elong2}) 
to the experimental points at 100 GeV using the
parameters calculated in Refs.\ \cite{bock81} and \cite{hughes90}. Note
that now we are not performing a fit but checking how well the general
form (\ref{elong2}) 
together with two sets of parameters for iron-scintillator
calorimeters describe our data. 
As shown in Fig.\ \ref{fig:f15}, both sets of
parameters work rather well in describing the 100 GeV data.

\begin{figure*}[tbph]
     \begin{center}
        \begin{tabular}{c}
\mbox{\epsfig{figure=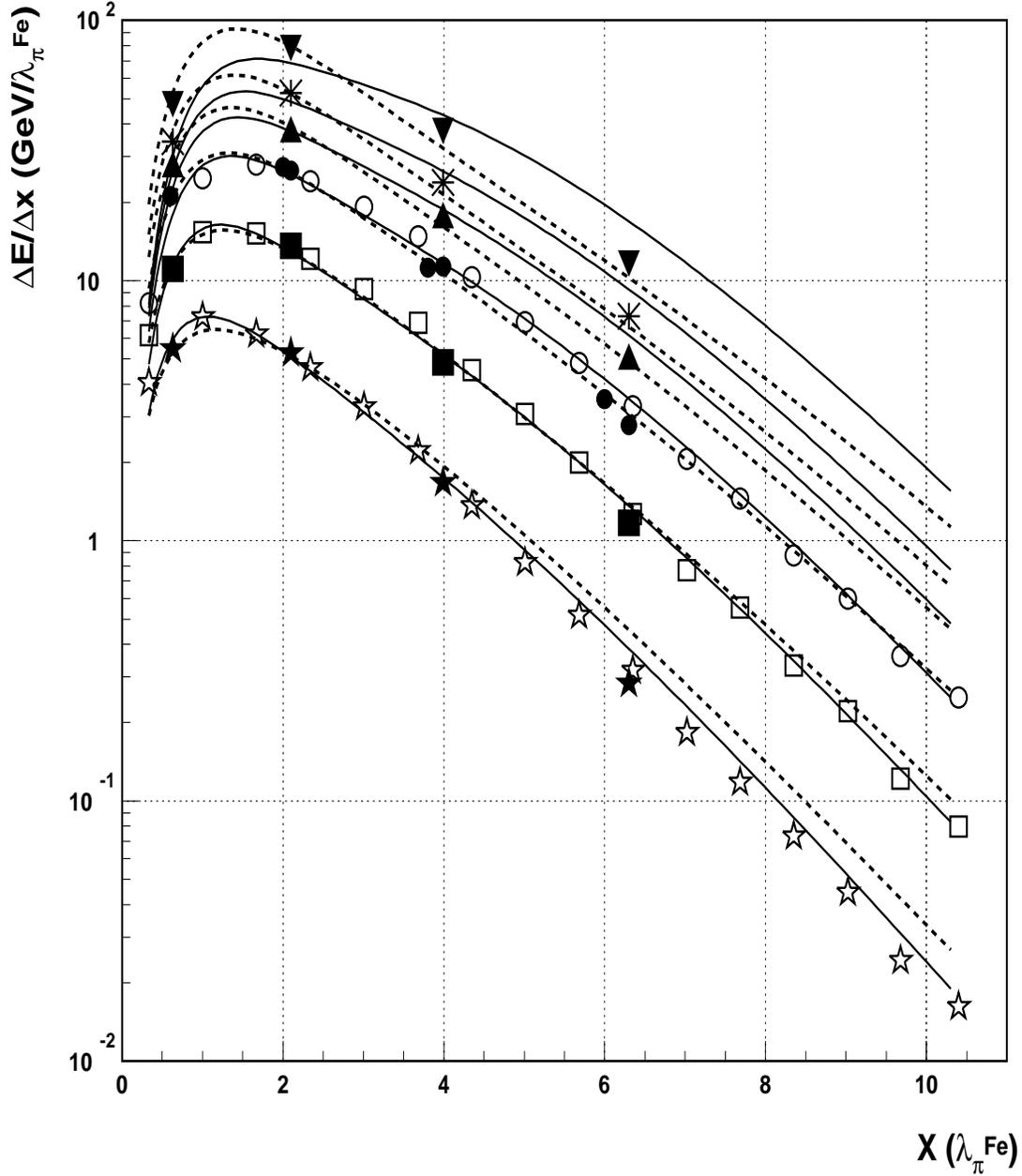,width=0.95\textwidth,height=0.75\textheight}} 
        \\
        \end{tabular}
     \end{center}
       \caption{
        Longitudinal profiles of the hadronic showers from 20 GeV (open 
        stars),
        50 GeV (open squares) and 100 GeV (open circles) 
        pions as a function of the longitudinal coordinate $x$ in units 
        of $\lambda_I$ for a conventional iron-scintillator 
        calorimeter \cite{hughes90} and of 20 GeV (black stars),
        50 GeV (black squares), 100 GeV (black circles), 
        150 GeV (black up triangles),
        200 GeV (asterisks), 300 GeV (black down triangles)
        for pions at $20^{\circ}$  
        and of 100 GeV (black circles) for pions at $10^{\circ}$
        for Tile iron-scintillator calorimeter. 
        The solid lines are calculated with function (\ref{elong2}) 
        with parameters from \cite{hughes90}.
        The dashed lines are calculated with function (\ref{elong2}) 
        with parameters from \cite{bock81}.
       \label{fig:f15-l}}
\end{figure*}

Turning next to the longitudinal shower development at different
energies, in Fig.\ \ref{fig:f15-l} our values of  $\Delta E / \Delta x$ 
for 20 -- 300 GeV together with the data from \cite{hughes90} are shown.
The solid and dashed lines are calculations with function (\ref{elong2})
using parameters from \cite{hughes90} and \cite{bock81}, respectively.
Again, we observe reasonable agreement between our data and the 
corresponding data for conventional iron-scintillator 
calorimeter on one hand,  and between data and the 
parametrizations described above.
Note that the fit in \cite{hughes90} has been performed in the energy 
range from 10 to 140 GeV; hence 
the curves for 200 and 300 GeV should be considered as extrapolations.
It is not too surprising that at these energies the agreement is
significantly worse, particularly at 300 GeV. 
In contrast, the parameters of \cite{bock81}
were derived from data spanning the range 15 -- 400 GeV, and are in much
closer agreement with our data.

\section{The parametrization of Hadronic Showers}

The three-dimensional parametrization for spatial hadronic shower
 development is
\begin{equation}
        \Psi (x, r) = 
                \frac{dE(x)}{dx} \cdot 
                \frac{ 
                \sum \limits_{i=1}^{3} \ \frac{a_{i} (x)}{\lambda_{i} (x)}\
                K_{0} \bigl( \frac{r}{\lambda_{i} (x)} \bigr)
                }{
                2\pi\ \sum \limits_{i=1}^{3} a_{i}(x) \lambda_{i} (x)} , 
\label{e123}
\end{equation}
where $dE  (x)/ dx$, defined by equation (\ref{elong2}),
is the longitudinal energy deposition,
the functions $a_{i}(x)$ and $\lambda_{i} (x)$ are given
by equations (\ref{eai}) and (\ref{eli}), and
$K_{0}$ is the modified Bessel function.

This explicit three dimensional parametrization can be used
as a convenient tool for many calorimetry problems requiring
the integration of a shower energy deposition in a volume
and the reconstruction of the shower coordinates. 

\section{Electromagnetic Fraction of Hadronic Showers}

One of the important issues in the understanding of hadronic showers
is the electromagnetic component of the shower, i.\ e.\ the fraction 
of energy going into $\pi^0$ production and its dependence on radial 
and longitudinal coordinates, $f_{\pi^{0}} (r, x)$.
Following \cite{acosta92},
we assume that the electromagnetic part of a hadronic shower is the
prominent central core, which in our case is the first term in the
expression (\ref{e23}) for the radial energy density function, $\Phi(r)$.
Integrating $f_{\pi^0}$ over $r$ we get
\begin{equation}
\label{e231}
        f_{\pi^{0}} = 
                \frac{a_1 \lambda_1}{\sum \limits_{i=1}^{3} a_i \lambda_i} .
\end{equation}
For the entire Tile calorimeter this value is $(53 \pm 3) \%$
at 100 GeV.

The observed $\pi^{0}$ fraction, $f_{\pi^{0}}$, is related to the 
intrinsic actual fraction, $f^{\prime}_{\pi^{0}}$, by the equation
\begin{equation}
        f_{\pi^{0}}(E) =
        \frac{e\ E^{\prime}_{em}}{e\ E^{\prime}_{em} + h\ E^{\prime}_{h}} = 
                \frac{ e/h \cdot f^{\prime}_{\pi^{0}}(E)}{(e/h - 1) 
                \cdot f^{\prime}_{\pi^{0}}(E) + 1} ,
\label{e506-1}
\end{equation}
where $E^{\prime}_{em}$ and $E^{\prime}_{h}$ are the intrinsic 
electromagnetic and hadronic parts of shower energy,
$e$ and $h$ are the coefficients of conversion of intrinsic electromagnetic 
and hadronic energies into observable signals, 
$f^{\prime}_{\pi^{0}} = E^{\prime}_{em}/(E^{\prime}_{em} + E^{\prime}_{h})$.

\begin{figure*}[tbph]
     \begin{center}
        \begin{tabular}{c}
\mbox{\epsfig{figure=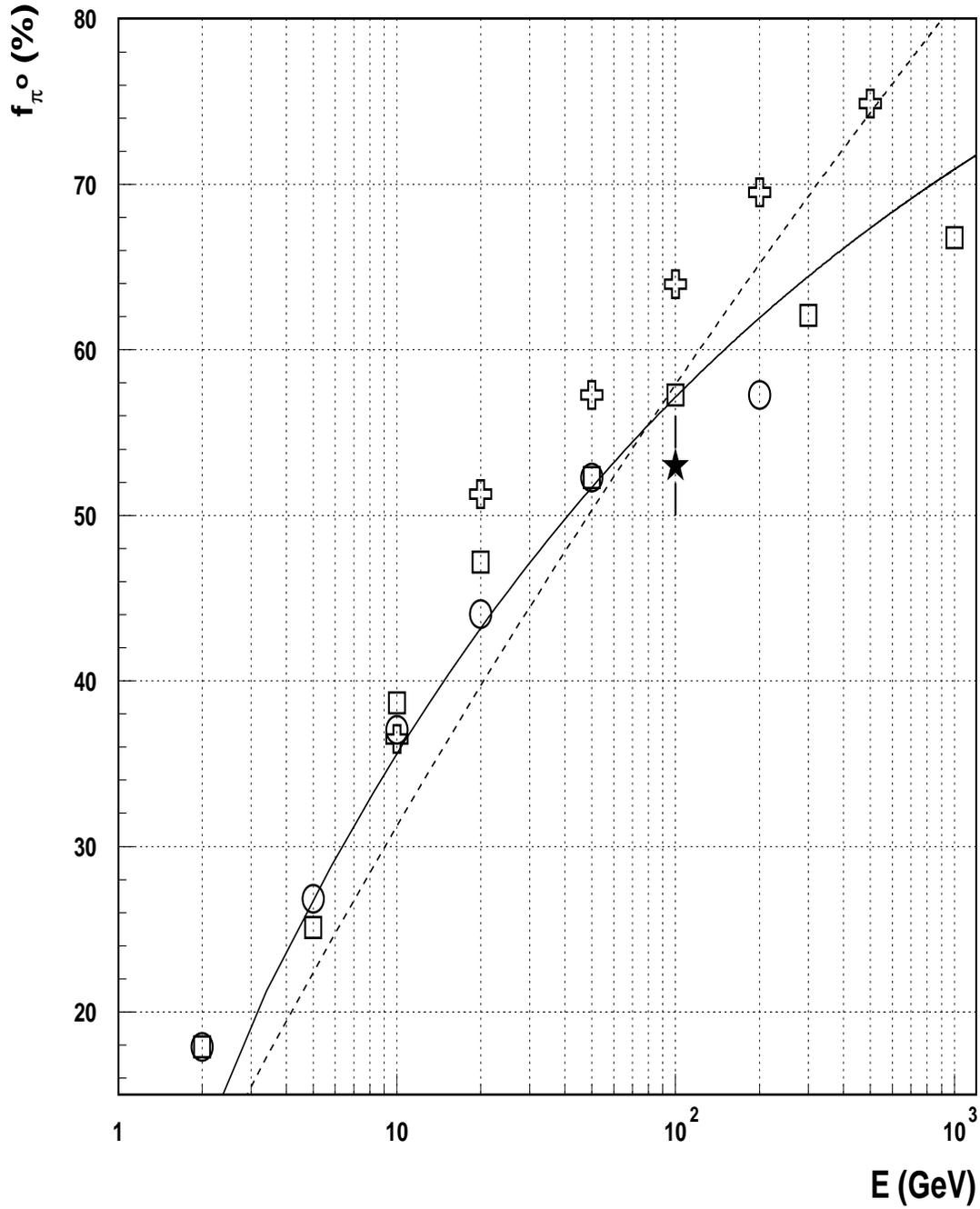,width=0.95\textwidth,height=0.80\textheight}} 
        \\
        \end{tabular}
     \end{center}
       \caption{
        The fraction $f_{\pi^{0}}$ in 
        hadronic showers versus the beam energy.
        The star is our data, 
        the solid curve is the Groom parametrization, 
        the dashed curve is the Wigmans parametrization,
        squares are the GEANT-CALOR predictions,
        circles are the GEANT-GHEISHA predictions and
        crosses are the CALOR predictions.
       \label{fig:f150-1}}
\end{figure*}

There are two analytic forms for the intrinsic $\pi^{0}$
fraction suggested by Groom
\cite{groom90}
\begin{equation}
        f^{\prime}_{\pi^{0}}(E) =
                1 - { \biggl( \frac{E}{E_{0}^{\prime}} \biggr) }^{m-1}
\label{e506-2}
\end{equation}
and Wigmans \cite{wigmans88}
\begin{equation}
        f^{\prime}_{\pi^{0}}(E) = 
                k \cdot ln \biggl( {\frac{E}{E_{0}^{\prime}}} \biggr),
\label{e506-3}
\end{equation}
where $E_{0}^{\prime} = 1$ GeV, $m = 0.85$ and $k = 0.11$.
We calculated $f_{\pi^{0}}$ using the value $e/h = 1.34\pm0.03$ 
for our calorimeter 
\cite{tilecal-tdr96}, \cite{budagov95-72}
and obtained the curves shown in Fig.\ \ref{fig:f150-1}.

Our result at 100 GeV is compared in 
Fig.\ \ref{fig:f150-1} 
to the modified Groom and Wigmans parametrizations and to results from
the Monte Carlo codes 
CALOR \cite{gabriel94}, 
GEANT-GEISHA \cite{juste95} and GEANT-CALOR \cite{bosman97} 
(the latter code is an implementation of
CALOR89 differing from GEANT-FLUKA only for 
hadronic interactions below 10 GeV). 
Note that the Monte Carlo calculations were performed for the intrinsic
$\pi^{0}$ fraction, $f^{\prime}_{\pi^{0}}(E)$, and therefore 
the results were modified by us according to (\ref{e506-1}).
As can be seen from Fig.\ \ref{fig:f150-1}, our calculated 
value of $f_{\pi^{0}}$ is about one standard deviation lower than 
two of the Monte Carlo results 
and the Groom and Wigmans parametrizations.

\begin{figure*}[tbph]
     \begin{center}
        \begin{tabular}{c}
\mbox{\epsfig{figure=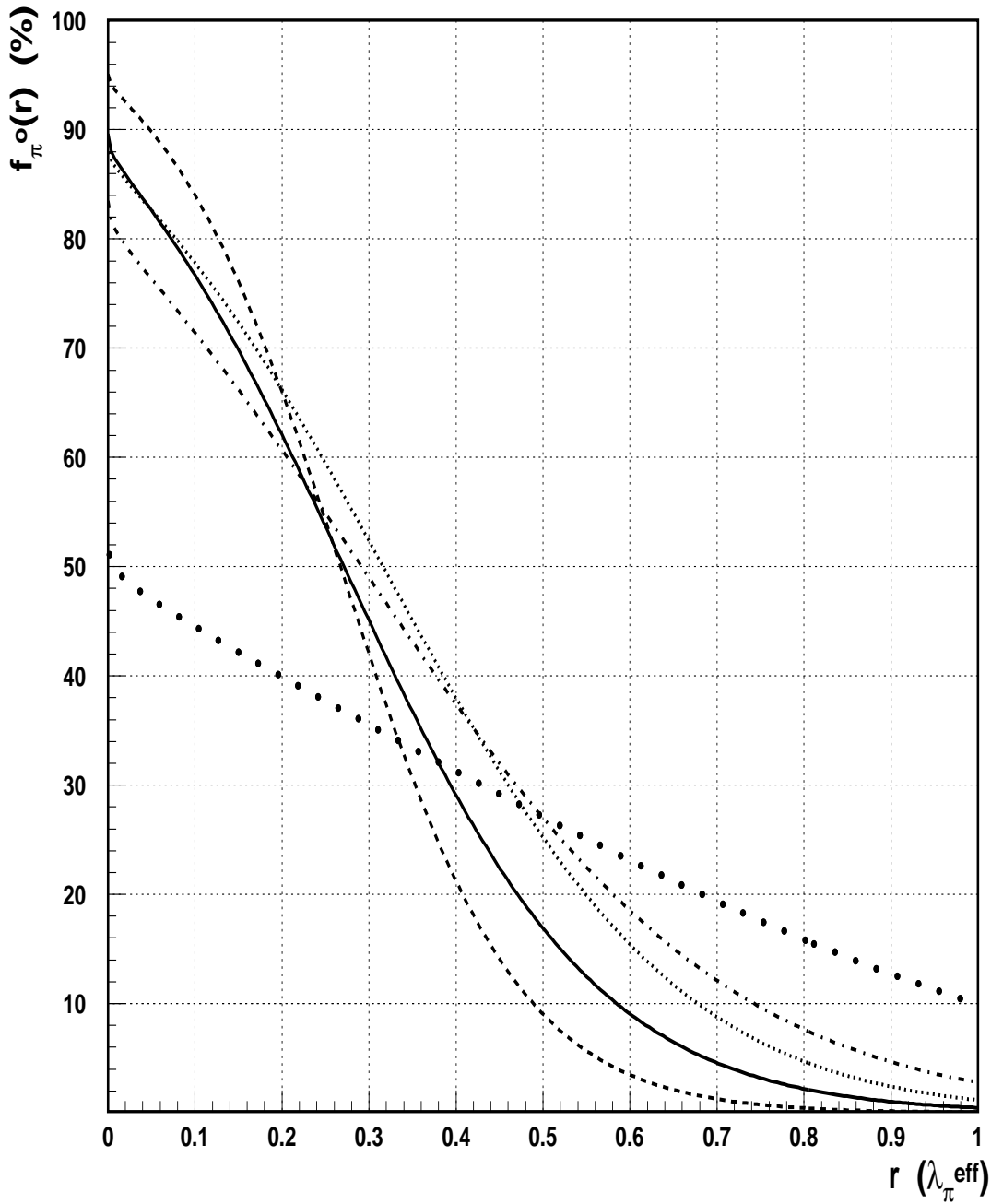,width=0.95\textwidth,height=0.80\textheight}}
        \\
        \end{tabular}
     \end{center}
       \caption{
        The $f_{\pi^{0}}$ fractions of hadronic showers 
        as a function of radius.
        The solid line is the $f_{\pi^{0}} (r)$ for the entire Tile 
        calorimeter.
        The dash-dotted line is the $f_{\pi^{0}} (r)$ for 
        the first depth segment,
        the dashed line is the $f_{\pi^{0}} (r)$ for 
        the second depth segment,
        the thin dotted line is the $f_{\pi^{0}} (r)$ for 
        the third depth segment,
        the thick dotted line is the $f_{\pi^{0}} (r)$ for 
        the fourth depth segment.
       \label{fig:f150-3}}
\end{figure*}

Figure \ref{fig:f150-3} shows the fractions $f_{\pi^{0}} (r)$ as a 
function of $r$.
As can be seen, the fractions $f_{\pi^{0}} (r)$ for the entire calorimeter 
and for depth segments 1 -- 3 amount to about $90\%$ as $r \rightarrow 0$ 
and 
decrease to about $1\%$ as $r \rightarrow \lambda_{\pi}^{eff}$.
However for depth segment 4 the value of $f_{\pi^{0}} (r)$ amounts 
to only $50\%$ as $r \rightarrow 0$ and decreases slowly to about $10\%$ as
$r \rightarrow \lambda_{\pi}^{eff}$. 

\begin{figure*}[tbph]
     \begin{center}
        \begin{tabular}{c}
\mbox{\epsfig{figure=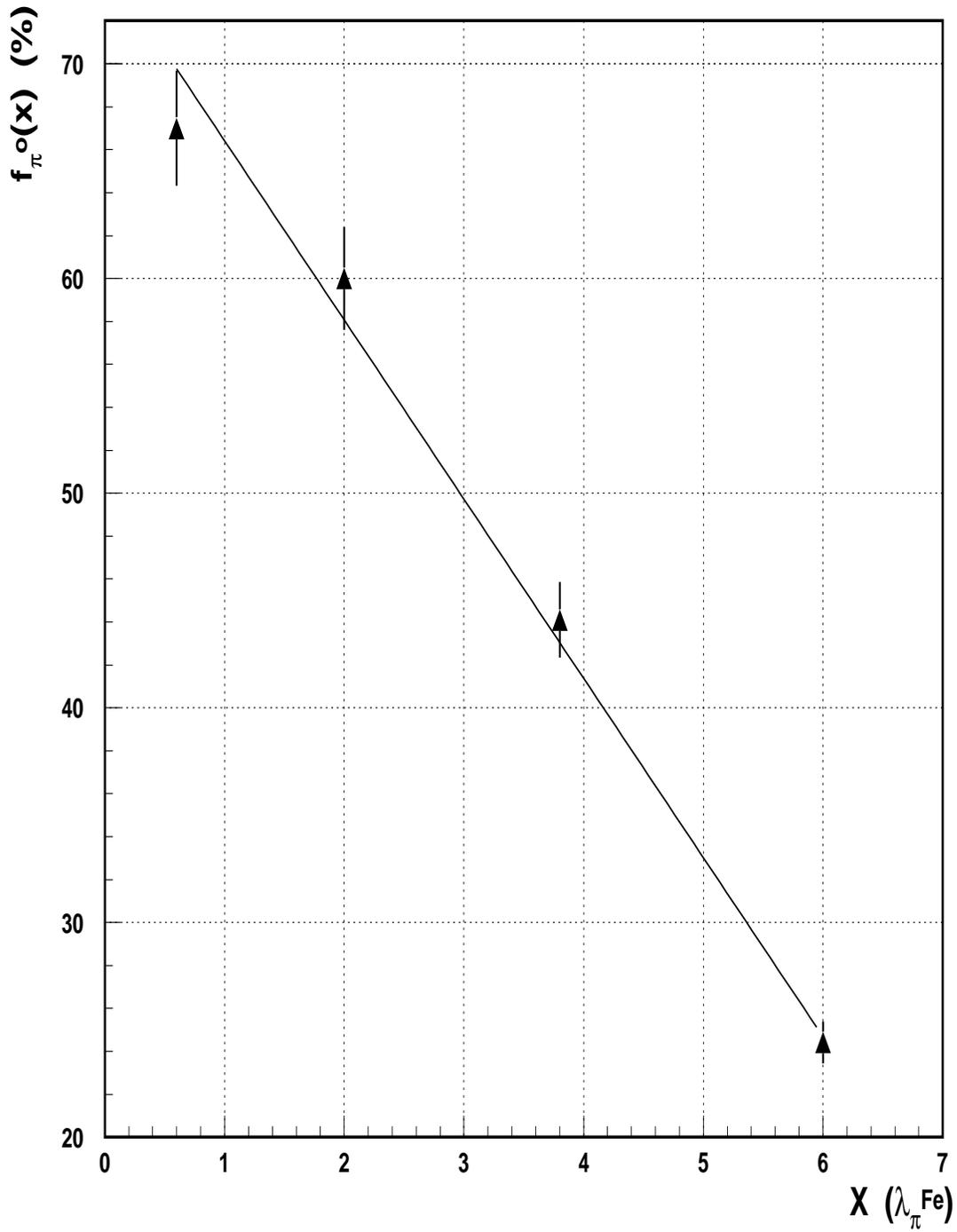,width=0.95\textwidth,height=0.85\textheight}}
        \\
        \end{tabular}
     \end{center}
       \caption{
        The $f_{\pi^{0}}(x)$ fractions of hadronic showers as a function 
        of $x$.
       \label{fig:f150-2}}
\end{figure*}

Figure\ \ref{fig:f150-2}
shows the values of $f_{\pi^{0}}(x)$  as a function of $x$,
as well as the linear fit which gives
$f_{\pi^{0}}(x) = (75\pm2) - (8.4\pm0.4) x$ ($\%$).

Using the values of $f_{\pi^{0}}(x)$ and energy depositions for various 
depth segments, we obtained the contributions from the electromagnetic and 
hadronic parts of hadronic showers in Fig.\ \ref{fig:f15}.
The curves represent a fit to the electromagnetic
and hadronic components of the shower using equation (\ref{elong00}).
$E_{f}$ is set equal to $f_{\pi^{0}} E_{beam}$
for the electromagnetic fraction and
$(1 - f_{\pi^{0}}) E_{beam}$ for the hadronic fraction.
The electromagnetic component of a hadronic shower rise and decrease 
more rapidly than the hadronic one 
($\alpha_{em} = 1.4\pm0.1$, $\alpha_{h} = 1.1\pm0.1$, 
$\beta_{em} = 1.12\pm0.04$, $\beta_{h} = 0.65\pm0.05$).
The shower maximum position
($x_{max} = (\alpha / \beta )\ \lambda_{\pi}^{eff}$)
occurs at a shorter distance from the calorimeter front face 
($x_{max}^{em} = 1.23\ \lambda_{\pi}^{eff}$, 
 $x_{max}^{h} = 1.85\ \lambda_{\pi}^{eff}$).
At depth segments greater than $4\ \lambda_{\pi}^{eff}$, the hadronic
fraction of the shower begins to dominate.
This is natural since 
the energy of the secondary hadrons is too low to permit significant 
pion production.

\section{Summary and Conclusions}

We have investigated the lateral development of hadronic showers
using 100 GeV pion beam data at an incidence angle of
$\Theta = 10^{\circ}$ for impact points $z$ in the range from 
$- 360$ to $200$ mm
and the longitudinal development of hadronic showers 
using 20 -- 300 GeV pion beams at an incidence angle of 
$\Theta = 20^{\circ}$.

Some useful formulae for the investigation of lateral profiles 
have been derived using a three-exponential form of
the marginal density function $f(z)$.

We have obtained for four depth segments and for the entire calorimeter:
energy depositions in towers, $E(z)$;
cumulative functions, $F(z)$;
underlying radial energy densities, $\Phi(r)$;
the contained fraction of a shower as a function of radius, $I(r)$;
the radii of cylinders for a given shower containment fraction;
the fractions of the electromagnetic and hadronic parts of a shower;
differential longitudinal energy deposition $\Delta E / \Delta x$; 
and a three-dimensional hadronic shower parametrization.

We have compared our data with those from a conventional
iron-scintilla\-tor ca\-lo\-ri\-me\-ter, those from a lead-scintillator
fiber calorimeter, and with Monte Carlo calculations.  
We have found that there is general reasonable agreement in the 
behaviour of the Tile calorimeter radial density functions
and those of the lead-scintillating fiber calorimeter;
that the longitudinal profile agrees with that of a conventional 
iron-scintillator calorimeter and the Monte Carlo predictions;       
that the value at 100 GeV
of the calculated
fraction of energy going into $\pi^{0}$
production in a hadronic shower, $f_{\pi^{0}}$, agrees with 
the Groom and Wigmans parametrizations and with 
some of the Monte Carlo predictions.

The three-dimensional parametrization of hadronic showers 
that we obtained allows direct use in any application that requires
volume integration of shower energy depositions  
and position reconstruction.
The experimental data on the transverse and longitudinal
profiles, the radial energy densities and the 
three-dimensional hadronic shower parametrization are 
useful for understanding the performance of the Tile calorimeter,
but might find broader application
in Monte Carlo modeling of hadronic showers, in particular in 
fast simulations, and for future calorimeter design.  

\section{Acknowledgements}

This paper is the result of the efforts of many people from the ATLAS
Collaboration.
The authors are greatly indebted to the entire Collaboration
for their test beam setup and data taking.
We are grateful to the staff of the SPS, and in
particular to Konrad Elsener, for the excellent beam conditions and
assistance provided during our tests.



\newpage

{

\vspace*{\fill}

\begin{table*}[tbh]
\caption{
        The parameters $a_i$ and $\lambda_i$ obtained by fitting 
        the transverse shower profiles for four depth segments and the entire 
        calorimeter at 100 GeV.
         \label{Tb2}}
\begin{center}
\begin{tabular}{@{}|c|c@{}|c|c@{}|c|c@{}|c|c|@{}}
\hline
          Depth
        & $x$ ($\lambda_{\pi}^{Fe}$) 
        & $a_1$ 
        & $\lambda_1$ (mm) 
        & $a_2$ 
        & $\lambda_2$ (mm) 
        & $a_3$ 
        & $\lambda_3$ (mm) 
\\
\hline
\hline
1
&0.6
&$0.88\pm0.07$ 
&$17\pm2$ 
&$0.12\pm0.07$ 
&$48\pm14$ 
&$0.004\pm0.002$ 
&$430\pm240$
\\
\hline
2
&2.0
&$0.79\pm0.06$ 
&$25\pm2$ 
&$0.20\pm0.06$ 
&$52\pm6$ 
&$0.014\pm0.006$ 
&$220\pm40$ 
\\
\hline
3
&3.8
&$0.69\pm0.03$ 
&$32\pm8$ 
&$0.28\pm0.03$ 
&$71\pm13$ 
&$0.029\pm0.005$ 
&$280\pm30$ 
\\
\hline
4
&6.0
&$0.41\pm0.05$ 
&$51\pm10$ 
&$0.52\pm0.06$ 
&$73\pm18$
&$0.07\pm0.03$ 
&$380\pm140$ 
\\ 
\hline
all four
&
&$0.78\pm0.08$ 
&$23\pm1$ 
&$0.20\pm0.08$ 
&$58\pm4$
&$0.015\pm0.004$ 
&$290\pm40$ 
\\
\hline
\end{tabular}
\end{center}
\end{table*}


\vspace*{\fill}

\begin{table*}[tbh]
\caption{
         The values of the parameters 
         $\alpha_i$, $\beta_i$, $\gamma_i$ and $\delta_i$.
         \label{Tb02}}
\begin{center}
\begin{tabular}{|c||c|c|||c||c|c|}
\hline
        
        & $\alpha_i$ 
        & $\beta_i$     ($1 / \lambda_{\pi}$)
        & 
        & $\gamma_i$    (mm) 
        & $\delta_i$    ($\mathrm{mm} / \lambda_{\pi}$) 
\\
\hline
\hline
        $a_1$
&       $0.99 \pm 0.06$ 
&       $- 0.088 \pm 0.015$ 
&       $\lambda_1$
&       $13 \pm 2$ 
&       $6 \pm 1$ 
\\
\hline
        $a_2$
&       $0.04 \pm 0.06$ 
&       $0.071 \pm 0.015$ 
&       $\lambda_2$
&       $42 \pm 10$ 
&       $6 \pm 4$ 
\\
\hline
        $a_3$
&       $- 0.001 \pm 0.002$ 
&       $0.008 \pm 0.002$ 
&       $\lambda_3$
&       $170 \pm 80$ 
&       $29 \pm 23$ 
\\
\hline
\end{tabular}
\end{center}
\end{table*}

\vspace*{\fill}

\newpage

\vspace*{\fill}

\begin{table*}[tbh]
\caption{
        The radii of cylinders for the given shower containment
        at 100 GeV.
        \label{Tb3}}
\begin{center}
\begin{tabular}{|c|c|c|c|}
\hline
$x$ ($\lambda_{\pi}^{Fe}$)&\multicolumn{3}{|c|}{$r$ ($\lambda_{\pi}^{eff}$)}\\
\cline{2-4}
         & 90$\%$ & 95$\%$ & 99$\%$ \\
\hline
\hline
0.6      & 0.44    & 0.64    & 1.43   \\
\hline
2.0      & 0.60    & 0.88    & 1.55   \\
\hline
3.8      & 0.92    & 1.27    & 1.75   \\
\hline
6.0      & 1.24    & 1.55    & 1.87   \\
\hline
all four  & 0.72    & 1.04    & 1.67   \\
\hline
\end{tabular}
\end{center}
\end{table*}

\vspace*{\fill}

\begin{table*}[tbh]
\caption{
        Average energy shower depositions at various depth segments
        at 100 GeV.
        \label{long-1}}
\begin{center}
\begin{tabular}{|c|c|c|c|}
\hline
 $x$                    ($\lambda_{\pi}^{Fe}$)
&$\Delta x$             ($\lambda_{\pi}^{Fe}$)
&$E_{0}$                  (GeV)
&$\Delta E / \Delta x$  (GeV/$\lambda_{\pi}^{Fe}$)      \\
\hline
\hline
0.6  & 1.2  & 25.0$\pm$0.3  & 20.8$\pm$0.3\\
\hline
2.0  & 1.6  & 42.8$\pm$0.2  & 26.8$\pm$0.1\\
\hline
3.8  & 2.0  & 22.0$\pm$0.1  & 11.0$\pm$0.1\\
\hline
6.0  & 2.4  &  8.4$\pm$0.5  &  3.4$\pm$0.2\\
\hline
\end{tabular}
\end{center}
\end{table*}

\vspace*{\fill}

}

\newpage


\clearpage
\newpage

\vspace*{\fill}
\vspace*{\fill}
\clearpage
\newpage

\clearpage
\newpage

\clearpage
\newpage

\vspace*{1cm}
\clearpage
\newpage

\clearpage
\newpage

\clearpage
\newpage

\clearpage
\newpage

\clearpage
\newpage

\clearpage
\newpage

\clearpage
\newpage

\clearpage
\newpage

\clearpage
\newpage

\clearpage
\newpage

\clearpage
\newpage

\clearpage
\newpage

\clearpage
\newpage

\clearpage
\newpage

\clearpage
\newpage

\clearpage

\end{document}